\documentclass[journal]{IEEEtran}

\usepackage{amsmath}
\usepackage{cite}
\usepackage{mathtools}

\usepackage{multicol}
\usepackage{soul}
\ifCLASSINFOpdf
\usepackage{graphicx}
\usepackage{multirow}
\usepackage{mathrsfs}
\usepackage{txfonts}
\else
\usepackage{graphicx}
\fi
\usepackage{float}
\usepackage{color}
\ifCLASSOPTIONcompsoc
\usepackage[caption=false,font=footnotesize,labelfont=sf,textfont=sf]{subfig}
\else
\usepackage[caption=false,font=footnotesize]{subfig}
\fi
\usepackage{cite}
\usepackage{dblfloatfix}
\usepackage{algorithm}
\usepackage{algpseudocode}
\usepackage{breqn}
\DeclareMathOperator*{\argmax}{arg\,max}
\usepackage{float}

\begin{document}
	\title{Reconfigurable and Intelligent Ultra-Wideband Angular Sensing: Prototype Design and Validation}
	
	\author{Himani~Joshi, Sumit~J.~Darak,~Mohammad Alaee-Kerahroodi and  Bhavani Shankar Mysore Rama Rao 
		\thanks{This work is supported by the funding received from CSIR, India under SRF Scheme along with IIIT Delhi ORF grants to Himani Joshi, Core research grant (CRG) awarded to Dr. Sumit J. Darak from DST-SERB, GoI and FNR Luxembourg BRIDGES project AWARDS.}
		\thanks{Himani Joshi and Sumit J. Darak are with Electronics and Communications Department,
			IIIT-Delhi, India-110020 (e-mail: \{himanij,sumit\}@iiitd.ac.in), Mohammad Alaee-Kerahroodi and  Bhavani Shankar Mysore Rama Rao are with Interdisciplinary Centre for Security, Reliability and Trust (SnT), University of Luxembourg, Luxembourg-1359 (e-mail: \{mohammad.alaee,Bhavani.Shankar\}@uni.lu)}
	}
	
	\maketitle
	
	\begin{abstract}
		The emergence of beyond-licensed spectrum sharing in FR1 (0.45-6 GHz) and FR2 (24 - 52 GHz) along with the multi-antenna narrow-beam based directional transmissions demand a wideband spectrum sensing in temporal as well as spatial domains. We referred to it as ultra-wideband angular spectrum sensing (UWAS), and it consists of digitization followed by characterization of the wideband spectrum. In this paper, we design and develop state-of-the-art UWAS prototype using USRPs and LabVIEW NXG for the validation in the real-radio environment. Since 5G is expected to co-exist with LTE, the transmitter generates the  multi-directional multi-user wideband traffic via LTE specific single carrier frequency division multiple access (SC-FDMA) approach. At the receiver, the first step of wideband spectrum digitization is accomplished using novel approach of integrating sparse antenna-array with reconfigurable sub-Nyquist sampling (SNS). The reconfigurable SNS allows the digitization of non-contiguous spectrum  via low-rate analog-to-digital converters, but it needs intelligence to choose the frequency bands for digitization. We explore multi-play multi-armed bandit based learning algorithm to embed intelligence. Compared to previous works, the proposed characterization (frequency band status and direction-of-arrival estimation) approach does not need prior knowledge of received signal distribution. The detailed experimental results for various spectrum statistics, power gains and antenna array arrangements along with lower complexity validate the functional correctness, superiority and feasibility of the proposed UWAS over state-of-the-art approaches.
		
	\end{abstract}
	
	
	\IEEEpeerreviewmaketitle
	
	\section{Introduction}
	
	Growing demand for spectrum, proliferation of heterogeneous services demanding a wide range of latency and power constraints, and exponential growth of wireless devices have led to the emergence of \textit{beyond-licensed} dynamic spectrum sharing \cite{nr1,nr2}. For instance, 5G networks are being deployed in an ultra-wide spectrum ranging from FR1 (0.45-6 GHz) and FR2 (24 - 52 GHz) comprising of licensed, shared as well as unlicensed spectrum \cite{nr1,nr2}. Furthermore, large propagation loss at high frequency demands multi-antenna narrow-beam based directive transmissions to achieve the desired quality-of-service. To enable dynamic spectrum access in such environments, ultra-wideband angular spectrum sensing (UWAS) needs to be investigated to identify the resources in temporal as well as spatial domains.
	
	UWAS comprises digitization of the wideband spectrum followed by the characterization of the desired bands of the spectrum. Since the wideband spectrum is sparsely occupied, sub-Nyquist sampling (SNS) using low rate analog-to-digital converters (ADCs) has been explored for direct digitization of the  wideband spectrum \cite{was01,was1,was2,was3,was4,tim_doa}. In SNS, the number of ADCs are directly proportional to the number of active transmissions in the wideband spectrum and hence leads to sensing failure (or characterization failure) whenever the number of active transmissions exceeds the number of ADCs.  To overcome this drawback, we recently proposed reconfigurable SNS \cite{was5,was6} which unlike the traditional SNS methods \cite{was01,was1,was2,was3,was4,tim_doa}, can dynamically digitize the non-contiguous bands of the wideband spectrum for a given number of ADCs and its architecture is based on non-contiguous SNS \cite{imp}. Reconfigurable SNS allows skipping of any number of bands in the wideband spectrum, and this is especially required when some part of the spectrum may not be useful due to high traffic, security constraints and high propagation loss. This enables the digitization of the spectrum significantly wider than that of the conventional SNS thereby leading to higher spectrum transmission opportunities (and hence, high network throughput) \cite{was5,was6}. Reconfigurable SNS demands additional intelligence to dynamically learn the spectrum statistics and select appropriate bands of the spectrum for digitization which can be accomplished using learning algorithms \cite{was5,was6}.

	The characterization of the narrowband spectrum, i.e. to estimate the parameter such as status (vacant/occupied), center frequency, modulation and direction-of-arrival (DoA) of the occupied band, interference, noise variance, has been discussed extensively in last decade \cite{was1,was2,was3,was4,was5,was6,interference1,interference2,amc}. Various state-of-the-art characterization algorithms have been studied and deployed in real networks. However, the characterization of the wideband spectrum is challenging since there is a need to reconstruct the spectrum from SNS samples before parameter estimation.
	
	The performance analysis of various UWAS approaches in the real-radio environment is a critical step towards their practical realization. However, there is limited work in this direction. For instance, the hardware prototypes of SNS in \cite{mwc_hw,wcl} are the state-of-the-art, but they consider wideband spectrum sensing only in the temporal domain. The extension of the temporal to spatial sensing demands multi-antenna transceivers which significantly increases the design complexity. Few prototypes to estimate the DoA of a user signal via multi-antenna receiver have been discussed in \cite{DoA1,DoA2,DoA3,DoA6,DoA7,DoA8,DoA9,DoA10}. All these works are focused on the narrowband spectrum and employ Nyquist-sampling based digitization. The main objective of the proposed work is to design and develop end-to-end prototype demonstrating reconfigurable and intelligent UWAS along with the experimental validation in the real-radio environment. In the proposed prototype, all baseband algorithms are realized using LabVIEW NXG, and Universal Software Radio Peripheral (USRPs) are used for communication over-the-air. 
	The contributions of the  paper are:
	\begin{enumerate}
		\item We develop a multi-antenna transmitter to generate the  multi-directional multi-user transmissions in a wideband spectrum. The multi-user signals are appropriately modulated via LTE specified single carrier frequency division multiple access (SC-FDMA) approach. The transmitter guarantees sparse occupancy in a wideband spectrum by allowing the frequency bands to switch between vacant and occupied states with certain probability distributions which is unknown at the receiver.
		
		\item At the receiver, we propose a novel wideband digitization approach via integration of the sparse antenna-array with reconfigurable SNS. We explore multi-play multi-armed bandit based learning algorithm to embed intelligence needed for dynamic selection of the non-contiguous spectrum in reconfigurable SNS. The proposed approach offers significant savings in receiver complexity over \cite{was01,was1,was4,was5} and feasible hardware architecture than \cite{was01,was2,was4,was6}.
		
		\item Compared to previous works, the proposed spectrum characterization (frequency band status and DoA estimation) approach does not need prior knowledge of received signal distribution and spectrum statistics.
		
		\item The detailed experimental results for various spectrum statistics, power gains and antenna array arrangements along with complexity analysis are presented to validate the functional correctness, superiority and feasibility of the proposed UWAS approach over various state-of-the-art UWAS approaches. 
		
	\end{enumerate}
	To the best of our knowledge, the proposed prototype is the first to integrate sparse antenna-array, reconfigurable and intelligent SNS along with characterization algorithms. Such prototype and subsequent experimental performance analysis of various UWAS approaches offer important insights which may not be possible in conventional simulation-based experiments. Other than prototype design, the proposed work also offers new contributions at the algorithm level compared to our previous works in \cite{was5, was6}. The proposed approach exploits sparse-array,  unlike the traditional uniform linear array (ULA) in \cite{was5} and offers a higher number of DoA estimations than \cite{was5}. While the proposed architecture is more hardware-amenable than \cite{was6} as later demands stringent control over delay elements in analog front-end (AFE). Please refer to Section II for more details.
	
	The rest of the paper is organized as follows. The related work in UWAS and hardware implementation is discussed in Section~\ref{related_work}, followed by the description of the system architecture of proposed reconfigurable and intelligent UWAS in Section~\ref{Sec3}. The design of the multi-user traffic generator and phase reference generator is described in Section~\ref{Sec4}. The design of the proposed UWAS receiver is explained in Section~\ref{Sec5}. The experimental performance, and hardware complexity are analyzed in Section~\ref{Sec6}, followed by the conclusions in Section~\ref{Sec7}.

	\vspace{-0.1cm}
	\section{Related Work}
	\label{related_work}
	In this section, a detailed review of works related to the wideband spectrum digitization and characterization approaches along with the prototypes for validation in the real-radio environment is presented.
	
	\vspace{-0.2cm}  
	\subsection{Wideband Digitization}
	In wideband digitization, we limit the discussion to multi-antenna SNS approaches as single-antenna SNS are not applicable for DoA estimation \cite{was01,was1,was2,was3,was4,was5,was6}. A nested array architecture with a delayed branch at every antenna of the dense array is proposed in \cite{was01}. Though it offers wideband spectrum digitization via low-rate ADCs, there are two major drawbacks: 1)~The delay in each branch must be equal, and it is of the order Nyquist period (i.e. in nano seconds), and 2) Each antenna needs two ADCs (one each at delayed and direct branch), and their analog bandwidth is equal to Nyquist rate, which is significantly wide. Both requirements are difficult to meet in AFE, making it difficult to realize in practice. The condition of exactly identical delay in each branch is relaxed  in \cite{was3}. Recently, compressed carrier and DoA estimation (CASCADE) architecture based on well-known modulated wideband converter (MWC) \cite{mwc} based SNS  has shown to overcome both drawbacks \cite{was1}. CASCADE uses $L$-shaped antenna array, and the AFE of each antenna consists of analog mixing function followed by low pass filter and ADC. Since mixer down-coverts each frequency band to the baseband, the analog bandwidth of ADC is reduced to the bandwidth of a frequency band. Later in \cite{tim_doa}, the estimation of signal parameters via rotational invariance technique (ESPIRIT) algorithm, used in the CASCADE method is modified to determine 2-D DoAs (i.e. angle of azimuth and elevation) along with the carrier frequency.  
	All these works \cite{was01,was1,was3,tim_doa} are based on contiguous SNS, which may not be suitable for next-generation wireless networks, as discussed in Section I. In \cite{was5}, we presented an intelligent reconfigurable SNS using ULA antenna arrangement with one antenna having multiple finite rate of innovation (FRI) branches of different mixing functions to overcome these two drawbacks.

	Ideally, digitization architecture should allow characterization of as many occupied bands in the digitized spectrum as possible. However, SNS restricts the number of allowable occupied bands in the digitized spectrum for a given number of antennas. For example, the number of occupied bands must be less than the number of antennas in \cite{was01,was1,was3,was5}. 
	The architectures in \cite{was2,was4,was6} allow characterization of the higher number of occupied bands than the number of antennas. To achieve this, \cite{was2} used uniform rectangular array {\color{black}having $L$ antennas} with one antenna containing $K$ multiple delayed branches along with 3-D spatial smoothing and rank enhancement based ESPIRIT algorithm. {\color{black} This allows \cite{was2} to sense up to $\frac{KL}{4}$ occupied bands.}
	In \cite{was4}, the architecture is based on the integration of ULA with multi-coset sampling \cite{mcs}. {\color{black} It sparsely activates only a few antennas of ULA to estimate DoA. But similar to \cite{was01,was1,was3}, both \cite{was2,was4} are also based on contiguous SNS. In \cite{was6}, we proposed a sparse antenna array and non-contiguous sampling based DoA estimation.} 
	However, due to the requirement of strict delay elements, \cite{was2,was4,was6} suffer from the same drawbacks as faced by \cite{was01}. {\color{black} Thus, there is a need for an architecture which allows non-contiguous SNS and support a higher number of DoA estimations.} Furthermore, as discussed in Section~I, prototype and validation of these approaches in the real-radio environment is an important task, and it has not been done yet in the literature.

	\begin{figure*}[b]
		\vspace{-0.35cm}
		\centering
		\includegraphics[scale=0.58]{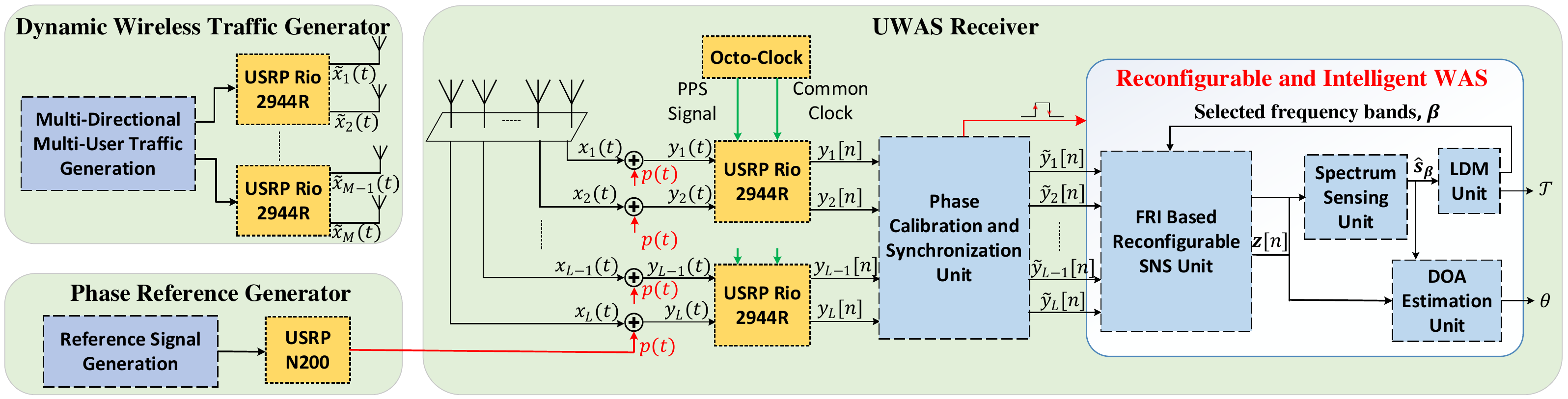}
		\caption{Proposed UWAS prototype consisting of 1) Dynamic Wireless traffic generator, 2) Phase reference generator, and 3) UWAS Receiver.}
		\label{BD}
	\end{figure*}
	
	\vspace{-0.2cm}
	\subsection{Spectrum Characterization}
	The spectrum characterization mainly involves the identification of the status (vacant/occupied) of each frequency band, estimation of modulation scheme, centre frequency, bandwidth and DoA of the occupied bands. In this paper, we focus on the identification of the status and DoA of occupied frequency bands.
	In \cite{mwc_hw,wcl,tim1,tim2,tim3,tim4,tim5,ss2,ss3}, various spectrum sensing methods are developed and validated on hardware testbeds. In \cite{tim1}, histogram based spectrum segmentation method is proposed to detect the spectral boundaries of primary (or licensed) users. Whereas in \cite{tim2,tim3,tim4}, the performance of various spectrum sensing methods like discriminant analysis \cite{tim2}, energy detector and  cyclo-stationary detector \cite{tim3} are validated. The performance of the cyclo-stationary detector for LTE SC-FDMA signal is analysed for different channel conditions in \cite{tim4}. The hardware set-up used in \cite{tim1,tim2,tim3,tim4} consists of a vector signal generator for the generation of a standardized primary user signal. Then via the Ethernet cable, this signal is then passed to the processing unit and vector signal analyzer to perform spectrum sensing. For validating the performance of energy detector under the non-ideal behaviour of receiver radio frequency (RF) AFE,
	\cite{tim5} uses software defined radio (SDR) for the reception of signal. But similar to \cite{tim1,tim2,tim3,tim4}, \cite{tim5} also performs wired transmission of the signal. To analyse the performance of spectrum sensing for over the air transmission, \cite{ss2,ss3} use USRPs for the transmission and reception of the signal. In \cite{ss2}, a cognitive radio network is developed to perform spectrum sensing, followed by dynamic spectrum access on the detected spectrum opportunities. Whereas a method identifying the selfish secondary (unlicensed) user, which does not contribute to collaborative spectrum sensing is proposed in \cite{ss3}. Although the spectrum sensing methods discussed in \cite{tim1,tim2,tim3,tim4,tim5,ss2,ss3} validate different spectrum sensing methods under various channels conditions and non-ideal behaviour of transmitter and receiver RF AFE, they all assume the Nyquist sampling based digitization. Hence, \cite{tim1,tim2,tim3,tim4,tim5,ss2,ss3} are not applicable to perform wideband spectrum sensing. 
	To overcome this drawback, \cite{mwc_hw} develops a prototype of MWC which implements contiguous SNS followed by the determination of the status of every frequency band. In \cite{wcl}, we discuss the non-contiguous spectrum sensing method along with its performance validation on the USRP testbed.

	The challenging part of the characterization is to estimate the DoA accurately as well as to optimize the number of maximum possible DoA estimations for a given number of ADCs and antennas.  In \cite{DoA1,DoA2}, the localization of a source signal under various hardware impairments like multi-path channel, non-ideal antenna array and non-ideal RF analog front end is discussed. To perform this task, single input multiple output testbed is developed for the DoA estimation. In \cite{DoA3}, DoA estimation at high altitude is accomplished via multi-antenna spherical array testbed. 
	In \cite{DoA6}, an interesting DoA estimation approach via a single antenna receiver on the high-speed train environment is demonstrated. By considering the uniform speed of the train, virtual ULA is realized by considering the samples received at different time instants along with space alternating generalized expectation-maximization (SAGE) principle \cite{DoA7} for DoA estimation.  For DoA estimation, MUSIC and ESPIRIT are widely used algorithms, and among them, MUSIC is shown to offer higher accuracy \cite{DoA8}.
	All these methods \cite{DoA1,DoA2,DoA3,DoA6,DoA7,DoA8} perform DoA estimation of only single user signal, i.e. narrowband DoA estimation. 
	
	Very few works consider the DoA estimation of the wideband spectrum and prototype design.
	The hardware testbeds discussed in \cite{DoA1,DoA2,DoA3,DoA6,DoA7,DoA8,DoA9} use uniform array for the DoA estimation and hence can not determine the DoAs when the number of users is more than the number of antennas. In \cite{DoA9}, DoA estimation testbed for two and three users with the help of four and eight antennas ULA, respectively is demonstrated. It employs Cholesky and LDL decomposition on the FPGA platform. \cite{DoA10} uses a co-prime antenna array to receive the signal and employs MUSIC, CAPON and LASSO methods to perform DoA estimation. Although \cite{DoA10} considers DoA estimation of multi-tone signal, all the tones have same DoA. 
	
	To summarize, existing works can perform the DoA estimation on Nyquist sampled signal and have limitations on the digitization bandwidth and the number of possible DoAs. Furthermore, none of the existing works offer an end-to-end prototype for UWAS.

	\vspace{-0.1cm}
	\section{Reconfigurable and Intelligent UWAS}
	\label{Sec3}
	Various building blocks of the proposed reconfigurable and intelligent UWAS are shown in Fig.~\ref{BD}. The prototype consists of three modules: 1) Dynamic Wireless traffic generator, 2) Phase reference generator, and 3) UWAS receiver.  The design details of each module are presented in the subsequent sections. The hardware units, i.e. USRPs and octo-clock, are shown using yellow-coloured blocks with a dotted border. USRPs are used for the transmission and reception of the wireless RF signals while the octo-clock is used for clock synchronization at the receiver. The blue-coloured blocks with dashed border  correspond to various signal processing, machine learning and wireless physical layer algorithms of the proposed UWAS and they are realized using LabVIEW NXG. Frequently used notations and symbols  are given in Table~\ref{tab_symbol}.
	
	Dynamic wireless traffic generator module emulates the multi-directional multi-user traffic in the wideband spectrum. For each user, we use LTE based SC-FDMA for baseband waveform modulation at the physical layer. The traffic of each user is beam-formed to a chosen direction via directional HORN-antennas integrated with USRPs. The transmit direction may change over time, distinct for each user and is chosen randomly. Since single NI-USRP 2944R has two transmitter/receiver ports, we need $M/2$ USRPs and $M$ directional antennas for the transmission of RF signal, $\tilde{x}_m(t)$ where $m\in\{1,2,...,M\}$, in $M$ directions. Furthermore, the carrier frequencies of users vary dynamically according to the probability distribution which maintains the sparsity of the wideband spectrum.
	We consider time-slotted communication which means that the carrier frequency and beam-direction of each user are constant over a given time slot, $t_s$ and may change dynamically from one slot to another.

	\begin{table}[htbp]
		\vspace{-0.2cm}
		\caption{Notations and their Definitions}
		\label{tab_symbol}
		\vspace{-0.2cm}
		\renewcommand{\arraystretch}{1.3}
		\resizebox{\linewidth}{!}{
			\begin{tabular}{l|l}
				\hline
				\textbf{Notation} & \textbf{Definitions}  \\
				\hline
				$M$ & Number of directional transmissions/users/antennas at transmitter\\
				\hline
				$L$ & Number of antennas of the phase antenna array at the receiver\\
				\hline
				$\tilde{\textbf{x}}(t)$ & $M$ multi-directional multi-user signals where $\tilde{x}_m(t)$ is the $m^{th}$ signal\\
				\hline
				$N$ & Number of multi-directional user data signal\\
				\hline
				\textbf{s} & Binary occupancy status of $N+1$ frequency bands where $s_i$ is the\\ &  status of $i\in\{0,1,\cdots,N\}$ frequency bands\\
				\hline
				$a_i(t)$ & Uncorrelated SC-FDMA signal transmitted in $i^{th}$ frequency band\\
				\hline
				$f_i$ & Carrier frequency of $i^{th}$ frequency band\\ 
				\hline
				$B$ & Bandwidth of a frequency band\\
				\hline
				$\mathcal{B}$ & Corresponds to the frequency range of $[0,B]$\\
				\hline
				$f_{prs}$ & Bandwidth of a phase reference signal\\
				\hline
				$f_r$ & Sum of the bandwidth of PRS and SS bands, i.e. $f_r = f_{prs}+B$ \\
				\hline
				$\beta$ & A set of non-contiguous frequency bands for UWAS\\
				\hline 
				$\textbf{s}_{\beta}$ & Contains binary occupancy status of only $\beta$ frequency bands\\
				\hline
				$\hat{\textbf{s}}_{\beta}$ & Estimated binary status of $\beta$ bands\\
				\hline
				$\textbf{p}_{uv}$ & A vector of transition probability from $u$ to $v$ binary states of all\\
				& $i\in\{1, 2,\cdots, N\}$ frequency bands\\
				\hline
				$f_t$  & Transmission/Reception frequency of the transmitter/receiver USRPs\\
				\hline
				$l$ & Receiver antennas index where $l\in\{1, 2, ... , L\}$\\
				\hline
				$\textbf{x}(t)$ & $L$ signals received by  phase antenna array where $x_l(t)$ is $l^{th}$ signal \\
				\hline
				$p(t)$ & Reference signal transmitted via SMA cable\\
				\hline
				$\textbf{y}(t)$ & $L$ outputs of the combiner i.e. $\textbf{x}(t) + p(t)$ where $y_l(t)$ is $l^{th}$ output \\
				\hline
				$\textbf{y}[n]$ & $L$ outputs of the receiver USRPs where $y_l[n]$ is the $l^{th}$ output signal\\
				\hline
				$\pmb{\phi}$ & Phase offset incurred in $L$ signals received from phase antenna array\\
				\hline
				$L_s$ & Length of the minimum sparse ruler of sparse antenna array\\
				\hline
				$\tilde{\textbf{y}}[n]$ & $L$ output signals of the phase calibration and synchronization unit \\
				& where $\tilde{y}_l[n]$ denotes the $l^{th}$ output signal\\
				\hline
				$\tilde{Y}_l(e^{j\omega})$ & Discrete time Fourier transform (DTFT) of $\tilde{\textbf{y}}[n]$ \\
				\hline
				$K$ & Number of additional branches in the FRI based reconfigurable SNS\\
				\hline
				$m[n]$ & Mixing function of FRI based reconfigurable SNS \\
				\hline
				$\gamma_{k,i}$ & Mixing coefficient of $m[n]$\\
				\hline
				$\tilde{z}_{k,l}[n]$ & $\{k,l\}^{th}$ output of the mixing function\\
				\hline
				$\tilde{Z}(e^{j\omega})$ & DTFT of $\tilde{z}_{k,l}[n]$ \\
				\hline
				$z_{k,l}[n]$ & $\{k,l\}^{th}$ output of the FRI based reconfigurable SNS \\
				\hline 
				$Z_{k,l}(e^{\omega})$ & DTFT of $z_{k,l}[n]$ \\
				\hline 
				$t_s$ & Denotes the current time slot\\
				\hline
				$\theta_i$ & DoA of the $i^{th}$ directional user signal received at the receiver\\ 
				\hline
				$\tau_l(\theta_i)$ & Time difference between the reception of $i^{th}$ user signal at the $l^{th}$ \\ & antenna and reference antenna \\
				\hline
				$\beta_{busy}$ & A set of occupied/busy frequency bands in $\beta$\\
				\hline
				$\textbf{E}$ & $L\times |\beta_{busy}|$ steering matrix containing $e^{j\omega_i\tau_l(\theta_i)}$ as its $\{l,i\}^{th}$ entry\\
				\hline
				$\zeta_{\beta}$ & Sensing failure event\\
				\hline
				$\psi(t_s)$ & Immediate probability of vacancy vector at time slot $t_s$\\
				\hline
				$T_s$ & Total time slot considered for the experiments\\
				\hline
				$\hat{\theta}_{g}$ & Estimated DoA at receiver antenna gain of $g~dB$\\
				\hline
				$\mathcal{T}$ & Total throughput achieved by a learning and decision making method\\
				\hline
				$\mathcal{R}$ & Total regret incurred by a learning and decision making method\\
				\hline
				$\theta_{err}$ & DoA estimation error\\
				\hline
				$\Delta$ & Maximum deviation in DoA estimation calculated for a fixed set-up\\
				\hline
				$\mathcal{F}(.)$ & Fourier transform operator\\
				\hline
				$vec(.)$ & Vectorization Operator\\
				\hline
				$\mathbb{P}(.)$ & Probability operator\\
				\hline
			\end{tabular}}
			\vspace{-0.75cm}
		\end{table}

		The UWAS receiver receives the multi-directional multi-user traffic signal, $x_l(t)$ where $l\in\{1,2,..,L\}$ and $L$ is the number of antennas at the receiver, via the designed sparse antenna array. Since the phase of the received signal is critical for accurate DoA estimation, phase reference generator module generates the reference signal, $p(t)$, which is combined with $x_l(t)$ and is responsible for performing phase calibration among the signals received at various antennas. 
		To digitize these $L$ combined signals, $y_l(t)$, we use $L/2$ NI-USRP 2944R. These USRPs receive the common clock signal and pulse per second (PPS) signal from the octo-clock unit to synchronize their local oscillators and ADCs, respectively. The signals, $y_l[n]$, received from the USRPs are digitized and downconverted to the desired sampling rate. 
		
		In baseband operation of UWAS, five tasks are performed in each $t_s$: 1)~Phase calibration and synchronization, 2)~FRI based reconfigurable SNS, 3)~Spectrum sensing, 4)~DoA estimation and, 5)~Learning and decision making (LDM) . Since UWAS requires the phase information of the signals impinging on the antenna array, the phase calibration of $y_l[n]$  removes the phase offset produced due to the independent RF channels of the receiver USRPs.	Subsequently, the filtering operation is performed to remove the reference signal, $p(t)$, and other synchronization signals to obtain the user data signal, $\tilde{y}_l[n]$ for subsequent digitization and characterization. The USRP performs digitization of the entire received signal at Nyquist rate. Hence, to perform non-contiguous UWAS over the selected frequency band, the desired frequency bands, $\beta$, are extracted from $\tilde{y}_l[n]$ via FRI based reconfigurable SNS.	The samples, $\textbf{z}[n]$, correspond to the sub-Nyquist samples of frequency bands present in $\beta$. $\textbf{z}[n]$ is passed to the spectrum sensing unit to determine the occupancy status, $\hat{\textbf{s}}_{\beta}$ of $\beta$ frequency bands, followed by the DoA estimation for the occupied frequency bands. At the same time, the learning and decision making unit updates the learned parameters and selects the $\beta$ frequency bands to be digitized in the subsequent time slot. 

		\section{Multi-User Traffic and Phase Reference Generation}
		\label{Sec4}
		In this section, we discuss the design details of dynamic wireless traffic and phase reference generator modules.
		\vspace{-0.15cm}
		\subsection{Dynamic Wireless Traffic Generator}
		\label{Sec4A}
\begin{figure}[!b]
			\vspace{-0.5cm}
			\centering
			\includegraphics[scale=0.55]{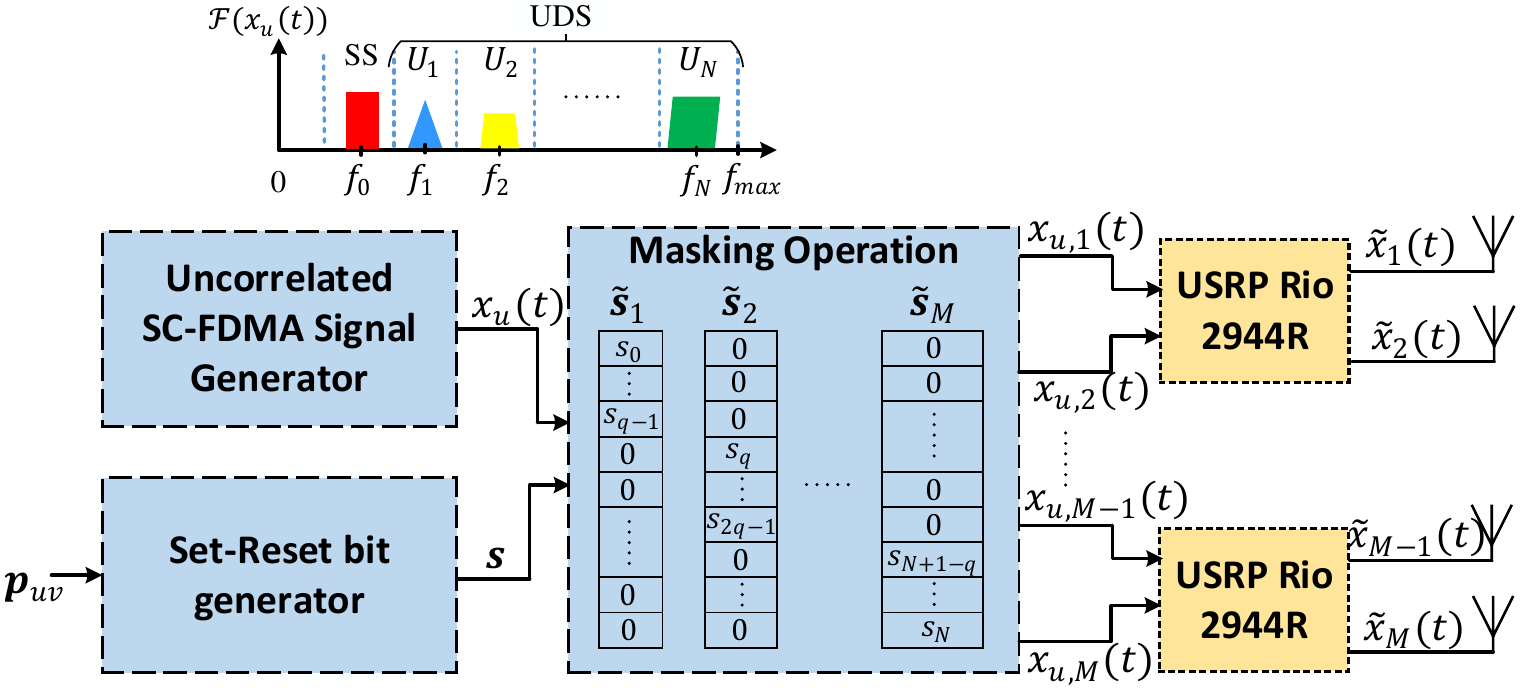}
			\caption{Dynamic wireless traffic generator.}
			\label{Tx}
			\vspace{-0.5cm}
		\end{figure}

		The dynamic wireless traffic generator, shown in Fig.~\ref{Tx}, consists of three sub-blocks. The first sub-block is the uncorrelated SC-FDMA signal generator. As shown in Fig.~\ref{Tx}, it generates a multiband signal, ${x}_u(t)$, which consists of $N+2$ frequency-bands, out of which one is reserved for the phase reference signal (more details are given in the next sub-Section) and another is reserved for the synchronization signal (SS). The remaining $N$ frequency bands (i.e. $U_1$ to $U_N$) are used for the user data communication and referred as user data signal (UDS). The SS, similar to the synchronization burst in the 4G/5G,  is used for frame and symbol synchronization over the downlink. The SC-FDMA signal generator block first generates the $N+1$ uncorrelated LTE SC-FDMA signals, $a_i(t)$ where $i\in\{0,1,...,N\}$, for SS and $U_1$ to $U_N$ frequency bands, and the modulates them to a carrier frequency of $f_i$.
		The bandwidth, $B$ of each SC-FDMA signal, i.e. each user can be varied between $1.4~MHz - 20 MHz$ similar to 4G. Mathematically,				\vspace{-0.25cm}
		\begin{equation}
			\vspace{-0.1cm}
			x_u(t) = \sum_{i = 0}^{N} a_i(t)e^{j2\pi f_i t}
		\end{equation}
		
		The second block is the set-reset bit generator block and it generates a binary status vector, $\textbf{s} = [s_0, s_1,..., s_N],$ consisting of masking bits for SS and $U_1$ to $U_N$ frequency bands. The occupancy of $U_1$ to $U_N$ frequency bands is decided based on the independent Markovian decision process (MDP). In MDP, the immediate occupancy status depends on the transition  probabilities of each of the $N$ frequency bands. Let $\textbf{p}_{uv}$ where $u,v\in\{0,1\}$ denotes the \{vacant,~busy\} status, be a vector storing the transition probabilities of $N$ frequency bands. Thus, $\textbf{p}_{uv}$ is an input to the second block.
		However, the status of these $N$ frequency bands changes only when the masking bit of the SS i.e. $s_0$ changes its status.
		To achieve this, the masking bit, $s_0$ is implemented as a square wave of 50\% duty cycle. So, whenever the status of $s_0$ changes, the masking bits $s_i~\forall~i\in\{1,...,N\}$ are updated according to the input $\textbf{p}_{uv}$.

		Since the designed multi-user traffic signal generates the directional traffic only in $M$ directions, the third block which performs masking operation, generates $M$ masking vectors, $\tilde{\textbf{s}}_{m}~\forall~m\in\{1, 2,..., M\}$ such that the output signals, ${x}_{u,m}(t)$ contain the information of bands $\{SS, U_1,.., U_{q-1}\}$, $\{U_{q}, .. , U_{2q-1}\}, ..... , \{U_{N+1-q}, ...., U_{N}\}$, where $q = \frac{N+1}{M}$. Thus, $\tilde{\textbf{s}}_{m} = [\textbf{0}_{1\times(m-1)q}, s_{(m-1)q}, ..., s_{mq-1}, \textbf{0}_{1\times(N+1-mq)}]$ where $\textbf{0}_{1\times q}$ denotes a  $1\times q$ size vector of zeros.
		Mathematically, the output signal, $x_{u,m}(t)$ of this block is written as
		\begin{equation}
			\vspace{-0.15cm}
			x_{u,m}(t) = \sum_{i=0}^N \tilde{\textbf{s}}_{m_i} a_i(t) e^{j2\pi f_i t}
		\end{equation}
		where $\tilde{\textbf{s}}_{m_i}$ denotes the $i^{th}$ entry of $\tilde{\textbf{s}}_{m}$.

		Now, the signals ${x}_{u,m}(t), ~\forall~m\in\{1,2,...,M\}$, are transmitted via $M$ channels of $\frac{M}{2}$ NI-USRP 2944R and $M$ HORN antennas. Please refer to Appendix~A for the actual LabVIEW NXG based implementation flowgraph of the dynamic wireless traffic generator.

		\vspace{-0.15cm}
		\subsection{Phase Reference Generator}
\begin{figure}[ht]
			\vspace{-0.15cm}
			\centering
			{\includegraphics[scale=0.7]{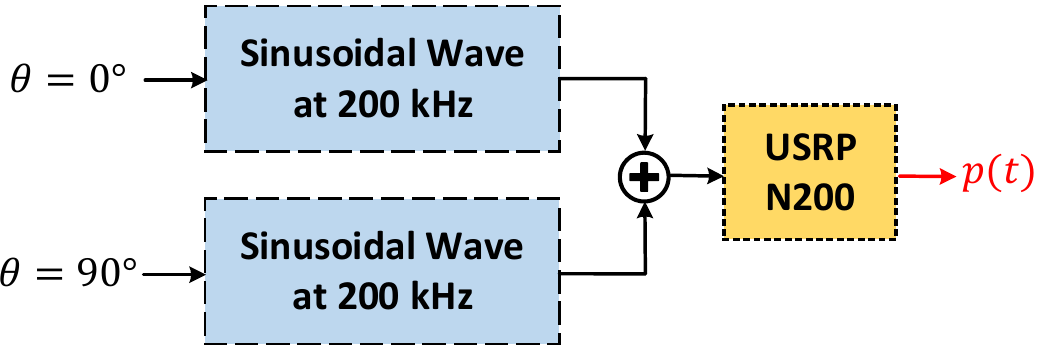}}
			\caption{Phase reference signal generator.}
			\label{Ref_bd}
			\vspace{-0.5cm}
		\end{figure}

		The UWAS receiver, shown in Fig.~\ref{BD}, receives the  multi-directional multi-user traffic signal, $x_l(t)$ where $l\in\{1,2..,L\}$  from a sparse antenna array of size $L$. 	The output of the antenna array is passed through independent AFE of the receiver USRPs for digitization. Since the AFE introduces phase distortion, we add a phase reference signal (PRS), $p(t)$, which can be used later to compensate for this distortion \cite{DoA_esti}. In our prototype, the signal received from the antenna is combined with PRS via SMA cables. PRS is generated entirely independent of the dynamic wireless traffic generator at the transmitter. Note that in existing 4G/5G system, there is a separate phase-tracking reference signal (PTRS) which is used to phase synchronize base-station and mobile terminals. As discussed in the next section, we also have similar signal and $p(t)$ is an additional signal to overcome the phase distortion of the receiver USRPs. The reference signal, $p(t)$ is based on the sinusoidal wave, as shown in Fig.~\ref{Ref_bd} where two sinusoidal signals  of carrier frequency $200~kHz$ and a phase shift of $0^o$ and $90^o$ are generated. These in-phase and quadrature-phase sinusoidal signals are then combined and passed through the USRP N200 for the baseband to RF conversion followed by interfaced with UWAS receiver via SMA cables.

		\section{UWAS Receiver}
		\label{Sec5}
		The UWAS receiver, shown in Fig.~\ref{BD}, receives the  multi-directional multi-user traffic signal, $\textbf{x}(t)$  from the antenna array of size $L$, which is then combined with $p(t)$. The resultant signal, $\textbf{y}(t)$, is passed through the NI-USRP 2944R for digitization and downconversion to obtain the multi-band signal, $\textbf{y}[n]$ as shown in Fig.~\ref{BD} and Fig.~\ref{Ref_Rx}. The signal $\textbf{y}[n]$ consists of three types of signals (also referred to as channels in 3GPPP 3G/4G standards): 1) Phase reference signal (PRS), $p(t)$  equivalent to PTRS in 4G/5G, 2) Synchronization signal (SS) equivalent to a primary synchronization signal (PSS) and secondary synchronization signal (SSS) in 4G/5G and 3) Multi-directional user-data signal (UDS) similar to physical downlink shared channel (PDSCH) in 4G/5G. 
		The distance between adjacent antennas of $L$-antenna array is carefully chosen to be integer multiple of $d=\frac{c}{2f_t}$, where $c$ is the speed of light and $f_t$ is the transmission frequency.
		This results in a sparse antenna array of length $L_s > L$  which in turn leads to a higher number of active DoA estimation for a given $L$. Please refer to Section~\ref{DoA_section} for more details. To characterize UDS, as shown in Fig.~1, the design of UWAS receiver consists of two units: 1) Phase calibration and synchronization unit, and 2) Reconfigurable and Intelligent WAS unit. Please refer to Appendix~B for the LabVIEW NXG based implementation flowgraph of the UWAS receiver.
		
		\subsection{Phase Calibration and Synchronization Unit}
		
\begin{figure}[ht]
			\vspace{-0.25cm}
			\centering
			{\includegraphics[scale = 0.58]{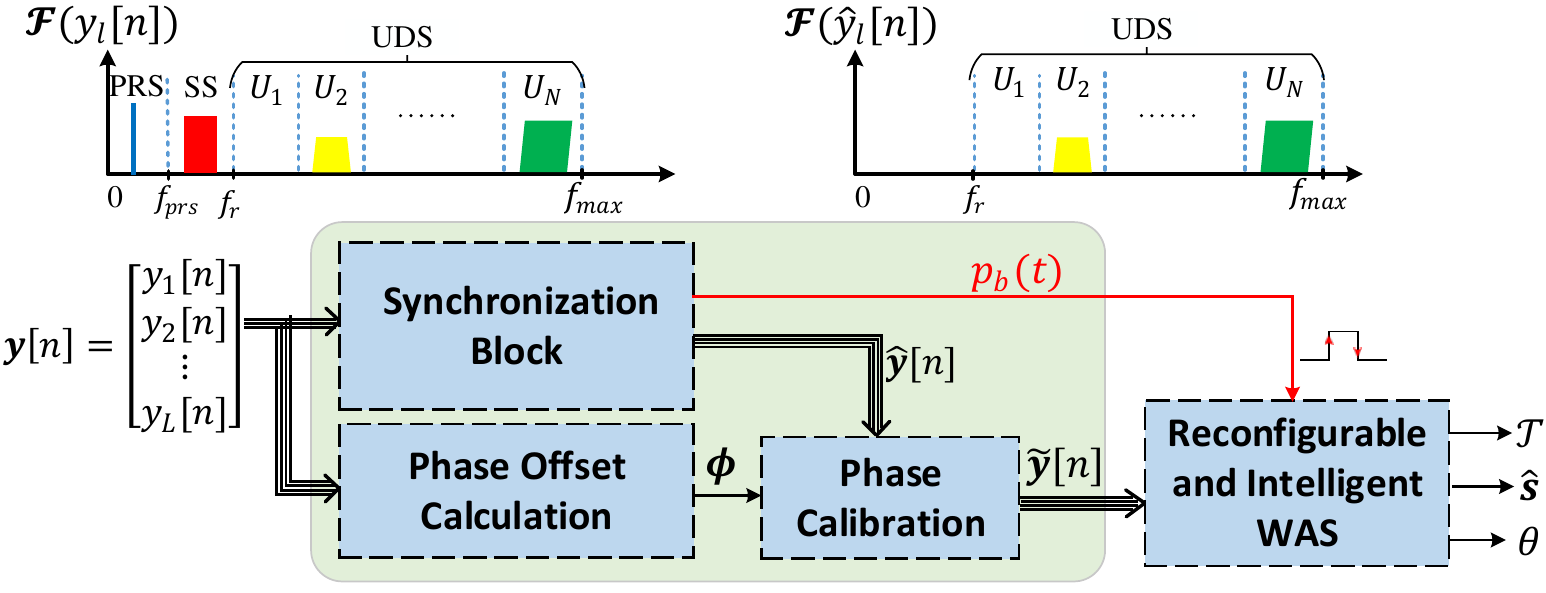}}
			\caption{Phase calibration and synchronization unit of the UWAS receiver.}
			\vspace{-0.5cm}
			\label{Ref_Rx}
		\end{figure}

		The phase calibration and synchronization unit, as explained in Fig.~\ref{Ref_Rx} consists of: 1)~Synchronization block, 2) Phase offset calculation block and 3) Phase calibration block. The first task of the synchronization block is to detect the SS signal and identify the slot boundary. Based on the slot boundary, it generates the pulse signal, $p_b(t)$ which is needed for WAS unit to differentiate between adjacent time slots and learn the spectrum statistics, i.e. $\textbf{p}_{uv}$. In the proposed prototype, we first filter the SS signal via band pass filter of cut-off frequencies $f_{prs}$ and $f_r = f_{prs} + B$ followed by an energy detection based approach to detect its status (vacant/occupied) and generate appropriate, $p_b(t)$. The second task of the synchronization block is to filter  $\textbf{y}[n]$ to obtain the UDS signal $\hat{\textbf{y}}[n]$ and forward it to the phase calibration block.

		
		The phase offset calculation block generates the phase offset, $\pmb{\phi}$, which is used by phase calibration block to eliminate the phase distortions caused by the different AFE of $L-$channels receiver USRPs. To determine $\pmb{\phi}$, the PRS, $p(t)$ is filtered out from ${\textbf{y}}[n]$ via low pass filter of cut-off frequency $f_{prs}$. The phase information of the filtered reference signal corresponds to the phase offset, $\pmb{\phi}$.
		


		Next, the phase calibration block receives the filtered UDS, $\hat{\textbf{y}}[n]$ and the phase offset, $\pmb{\phi}$ to perform phase calibration on $\hat{\textbf{y}}[n]$. Here, the contribution of phase offset is removed from $\hat{\textbf{y}}[n]$. For the $l^{th}$ signal, the phase calibration is performed as
		\begin{equation}
			\tilde{y}_l[n] = \hat{y}_l[n] e^{-j\phi_l}
		\end{equation}
		
		Hence, the phase of $\tilde{\textbf{y}}[n]$ only contains the phase information introduced by the sparse antenna array. The signal, $\tilde{\textbf{y}}[n]$, is then passed to the reconfigurable and intelligent WAS block for subsequent baseband processing and learning tasks.

		\begin{figure}[t]
			\vspace{-0.25cm}
			\centering
			\includegraphics[scale=0.68]{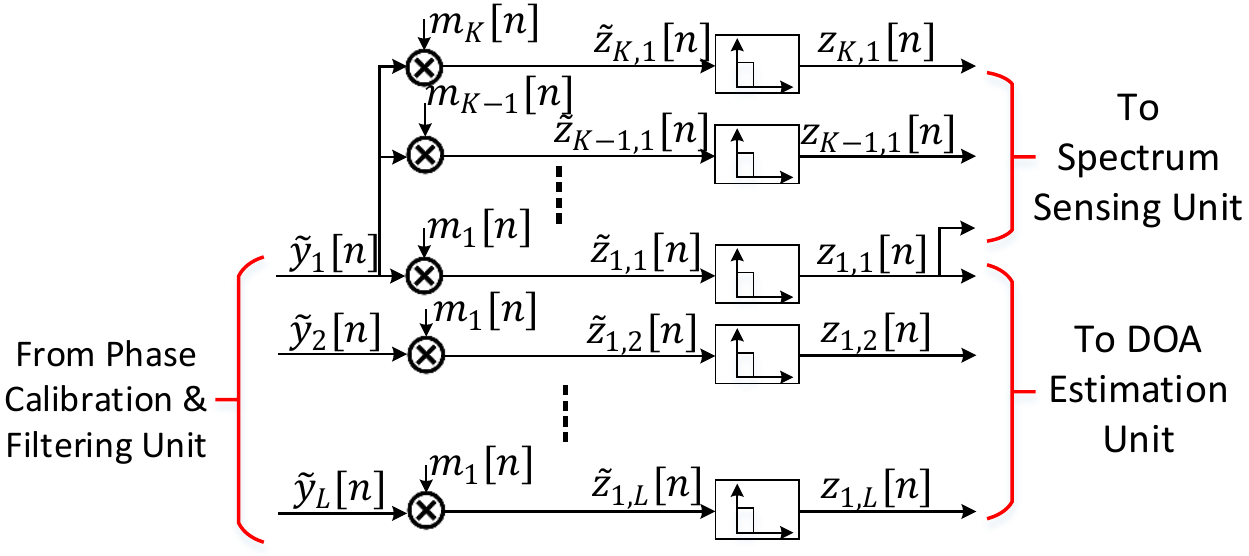}
			\caption{FRI based reconfigurable SNS.}
			\label{fri}
			\vspace{-0.25cm}
		\end{figure}

		\subsection{Reconfigurable and Intelligent WAS Unit}
		The reconfigurable and intelligent WAS is the brain of the UWAS receiver, which performs parameter estimation and learning and decision making (LDM) tasks. The parameter estimation includes status (vacant/occupied), $\textbf{s}_{\beta}$, of digitized frequency bands, $\beta$, and DoA, ${\theta}$, of the occupied frequency bands in $\beta$. The LDM includes learning the spectrum statistics of different frequency bands in the wideband spectrum and 
		selecting  the frequency bands to be digitized in the subsequent time slot, so the total throughput, $\mathcal{T}$ (i.e. transmission opportunities) is increased. 
		Instead of digitizing all frequency bands, the SNS approach realized in our prototype allows the dynamic digitization of frequency bands selected in a given time slot. Thus, we refer it to as reconfigurable SNS. For reconfigurable SNS, we embed intelligence to learn and identify the frequency bands to be digitized so as to improve throughput, i.e. maximize the number of vacant bands among the frequency bands selected for digitization.

		The signal, $\tilde{\textbf{y}}[n]$, received from the phase calibration and synchronization unit comprises of the samples of the wideband spectrum. To realize the reconfigurable SNS, the  signal $\tilde{\textbf{y}}[n]$ is passed through FRI based SNS architecture consisting of $L$ AFE, as shown in Fig.~\ref{fri} \cite{was5}.
		Mathematically, the wideband signal consisting of transmission from multiple users at the $l^{th}$ AFE can be represented as
		\begin{equation}
			\tilde{y}_l[n] = \sum_{i=1}^N s_i(t_s) a_i[n]e^{j\omega_i (n+\tau_l(\theta_i))} + \eta_l[n]
		\end{equation}
		where $s_i(t_s)\in \textbf{s}$ is the transmission/occupancy status of frequency bands at a time instant $t_s$ with $s_i(t_s) = 0$ denotes the no transmission (i.e. $i^{th}$ band is vacant) and $s_i(t_s) = 1$ denotes the transmission of active SC-FDMA signal. Please note that as discussed in Section~\ref{Sec4A}, $s_i(t_s)$ is unknown at the receiver and for the simplicity of notations, $s_i(t_s)$ is denoted as $s_i$. $a_i[n]$ is a discrete time SC-FDMA signal transmitted at the $i^{th}$ frequency band of a center frequency, $\omega_i=2\pi f_i$, $\theta_i$ is the direction of arrival of $a_i[n]$, $\tau_l(\theta_i)$ is the time difference between the reception of signal, $a_i[n]$ at the $l^{th}$ antenna and the reference antenna and $\eta_{l}[n]$ is the additive white Gaussian noise at $l^{th}$ received signal. Please note that $\tau_l(\theta_i)$ is dependent on $\theta_i$ and the antenna array structure. The discrete time Fourier transform (DTFT) of $\tilde{y}_l[n]$ i.e. $\mathcal{F}(\tilde{y}_l[n])$, is given as
		\begin{equation}
			\tilde{Y}_l(e^{j\omega}) = \sum_{i=1}^N s_i e^{j\omega_i \tau_l(\theta_i)} A_i(e^{j(\omega-\omega_i)}) + \eta_{l}(e^{j\omega})
		\end{equation}
		where $A_i(e^{j\omega})$ is the DTFT of $a_i[n]$. To generate samples corresponding to a set of specific frequency bands (of indices are stored in $\beta$), $\tilde{y}_l[n]~\forall~l\in\{1,..,L\}$, is passed through a mixing unit. The mixing function, $m_k[n]$ is defined as
		\begin{equation}
			m_k[n] = \sum_{b\in \beta}\gamma_{k,b}e^{-j2\pi((b-1)B+f_r)n}
		\end{equation}  
		where $\gamma_{k,b}$ is a mixing coefficient and is generated randomly from Gaussian distribution, $f_r$ is the frequency offset due to PRS and SS, and $B$ is the bandwidth of a frequency band or a SC-FDMA signal. Under noiseless condition, the DTFT of the output of the mixing unit is
		\begin{align} 
			\tilde{Z}_{k,l}(e^{j\omega}) 
			&= \sum_{n=-\infty}^{+\infty} \sum_{i=1}^N s_i a_i[n]e^{j\omega_i (n+\tau_l(\theta_i))} \sum_{b\in \beta}\gamma_{k,b}e^{-j2\pi((b-1)B+f_r)n}~e^{-j\omega n}\\
			&= \sum_{i=1}^N  e^{j\omega_i\tau_l(\theta_i)} s_i \sum_{b\in \beta}\gamma_{k,b} \sum_{n=-\infty}^{+\infty} a_i[n]			e^{-j2\pi(f-(f_i-(b-1)B-f_r))n}\\
			& = \sum_{i=1}^N  e^{j\omega_i\tau_l(\theta_i)} s_i \sum_{b\in \beta}\gamma_{k,b} A_i(e^{j2\pi(f-(f_i-(b-1)B-f_r))})
		\end{align}
		
		\begin{figure}[!b]
			\centering
			\vspace{-0.35cm}	\includegraphics[scale=0.62]{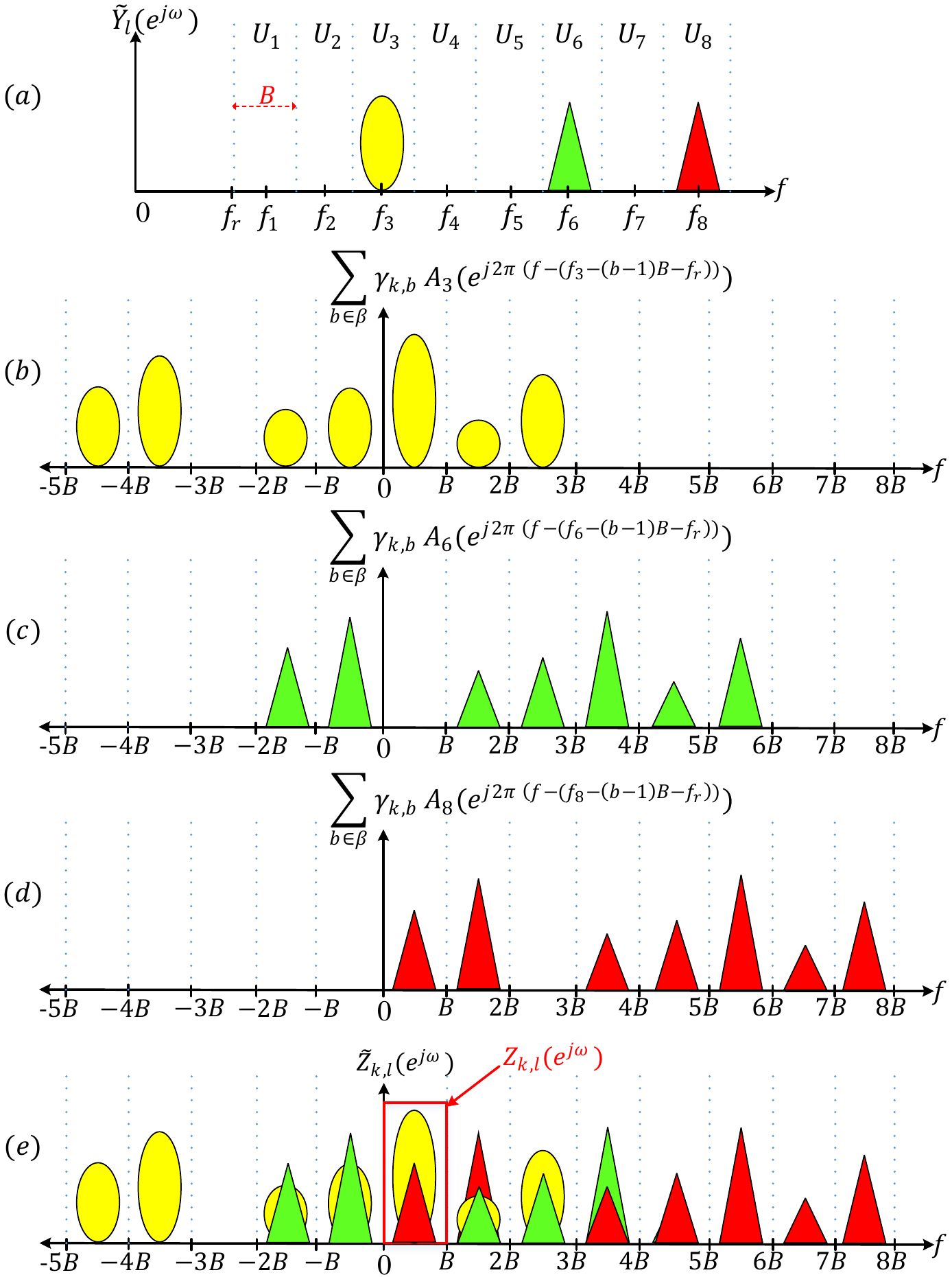}
			\caption{ 
				DTFT of the (a) $l^{th}$ output of Phase  calibration  and  synchronization unit i.e. $\tilde{y}_l[n]$, (b) $\tilde{z}_{k,l}[n]$ for $i = 3$ in Eq.~\ref{Z_tilde}, (c) $\tilde{z}_{k,l}[n]$ for $i = 6$ in Eq.~\ref{Z_tilde}, (d)
				$\tilde{z}_{k,l}$ for $i = 8$ in Eq.~\ref{Z_tilde} and (e) $\tilde{z}_{k,l}[n]$ and output of LPF ${z}_{k,l}[n]$ (in the red box)}
			\label{ex}
		\end{figure}
		
		For illustration, consider a scenario as shown in Fig.~\ref{ex}(a) where $N=8$ and at a given time slot, $t_s$, $\textbf{s} = \{0,0,1,0,0,1,0,1\}$ and $\beta = \{1,2,3,4,5,7,8\}$. Let 
		$\mathcal{N}_{busy}$ be a set containing the indices of occupied frequency band in $\tilde{\textbf{y}}[n]$ and $\beta_{busy}$ be a set containing the indices of occupied bands in $\beta$. This means $\mathcal{N}_{busy} = \{3,6,8\}$ and $\beta_{busy}=\{3,8\}$. Thus, as discussed in the Eq.~9, $\tilde{Z}_{k,l}(e^{j\omega})$ will 
		contain the contribution of only those bands for which $s_i = 1$ where $s_i\in\textbf{s}$, i.e. for $\mathcal{N}_{busy}$  frequency bands.
		Therefore, Eq.~9 is written as
		\begin{equation}
			\vspace{-0.1cm}
			\tilde{Z}_{k,l}(e^{j\omega}) = \sum_{i\in \mathcal{N}_{busy}}  e^{j\omega_i\tau_l(\theta_i)} \sum_{b\in \beta}\gamma_{k,b} A_i(e^{j2\pi(f-(f_i-(b-1)B-f_r))})
			\label{Z_tilde}
		\end{equation}
		Fig.~\ref{ex}(b)-(d) show Eq.~\ref{Z_tilde} for all $i\in \mathcal{N}_{busy}$ (i.e. $i \in \{3,6,8\}$) and it can be  observed that only for $\beta_{busy}$ bands i.e. $U_3$ and $U_8$, $A_i$ is present in the frequency range $\mathcal{B}=[0,B]$. However, it is noticed from Fig.~\ref{ex}(e) that Eq.~\ref{Z_tilde} contains images outside $\mathcal{B}$. Thus, after applying LPF on $\tilde{Z}_{k,l}(e^{j\omega})$ over $\mathcal{B}$, the DTFT of the output is
		\begin{equation}
			\vspace{-0.1cm}
			Z_{k,l}(e^{j\omega}) = \sum_{i\in \beta_{busy}} e^{j\omega_i\tau_l(\theta_i)} \gamma_{k,i}~A_i(e^{j\omega})
			\label{final_eq}
		\end{equation}
		Now, to perform WAS, the samples $z_{k,1}[n]~\forall~k\in\{1,...,K\}$ are passed to the spectrum sensing unit to determine the estimated status, $\hat{\textbf{s}}_\beta$ of $\beta$ frequency bands selected by LDM algorithm and digitized by reconfigurable SNS. Further, the samples $z_{1,l}~\forall~l\in\{1,..,L\}$ are passed to the DoA estimation unit to estimate the DoA of detected busy bands in $\beta$.   

		\subsubsection{Spectrum Sensing Unit}
		The aim of spectrum sensing (SS) unit is to estimate the status,  ${\textbf{s}}_{\beta}$, of $\beta$ frequency bands. Since SS uses the output, $z_{k,l}[n]$ where $k\in\{1,...,K\}$ and $l=1$, Eq.~\ref{final_eq} can be represented as
		\begin{align}
			Z_{k,1}(e^{j\omega}) &= \sum_{i\in \beta_{busy}} \gamma_{k,i} C_i(e^{j\omega})
			\equiv \sum_{i\in\beta} \gamma_{k,i} C_i(e^{j\omega})
			\label{SS}
		\end{align} 
		where $C_i(e^{j\omega}) = e^{j\omega_i\tau_1(\theta_i)} ~A_i(e^{j\omega})~\forall~i\in \beta_{busy}$ and is $0$ otherwise. For all values of $k$, Eq.~\ref{SS} can be written in matrix form as
		\begin{equation}
			\textbf{Z}_{\textbf{k}} = \pmb{\gamma} \textbf{C}
			\label{SS2}
		\end{equation}
		where $\pmb{\gamma}$ is a $K\times |\beta|$ matrix with $\gamma_{k,i}$ as its $\{k,i\}^{th}$ entry and $\textbf{C}$ is a sparse matrix with $|\beta_{busy}|$ non-zero rows. Thus, the estimation of ${\textbf{s}}_{\beta}$ from Eq.~\ref{SS2} can be treated as compressive sensing (CS) problem. Several CS algorithms have been studied in the literature \cite{sparse_rec}. These algorithms are mainly categorized as 1) greedy algorithm like orthogonal matching pursuit (OMP), Bayesian matching pursuit (BMP) algorithm, etc. and 2) $l_1$ minimization like basis pursuit (BP), least absolute shrinkage and selection operator (LASSO), etc. The $l_1$ minimization algorithms offer better reconstruction accuracy and do not require any prior knowledge of spectrum sparsity. However, these algorithms have higher computational time than the greedy algorithm. 
		The LDM learns the spectrum statistics, i.e. the prior probability, $\textbf{p}_{uv}$, which makes BMP algorithm best fit for the determination of $\textbf{C}$.
		But BMP also requires the knowledge of probability distribution function of $P(\textbf{C}|\textbf{s})$, and as we are transmitting SC-FDMA signal in the multi-user traffic, this parameter is unknown for the considered signal model. Hence, we apply the OMP algorithm \cite{omp} to determine the estimated status, $\hat{\textbf{s}}_{\beta}$ of $\beta$ frequency bands. 
		
		\subsubsection{DoA Estimation Unit}
		\label{DoA_section}
		This unit aims to estimate the DoA of detected busy frequency bands i.e. $\beta_{busy}$. To perform this task, DoA unit utilizes the estimated status, $\hat{\textbf{s}}_\beta$ and the samples $z_{1,l}~\forall~l\in\{1,...,L\} $. For the DoA estimation, Eq.~\ref{final_eq} can be rewritten as
		\begin{equation}
			Z_{1,l}(e^{j\omega}) = \sum_{i\in \beta_{busy}} e^{j\omega_i\tau_l(\theta_i)} D_i(e^{j\omega})
			\label{DoA1}
		\end{equation} 
		where $ D_i(e^{j\omega}) = \gamma_{1,i}~A_i(e^{j\omega})$. All $l\in\{1,...,L\}$ can be represented in the matrix form as
		\begin{equation}
			\textbf{Z}_{l} = \textbf{E}~\textbf{D}
			\label{DoA_general}
		\end{equation}
		where $\textbf{E}$ is a $ L\times |\beta_{busy}|$ steering matrix with $e^{j\omega_i\tau_l(\theta_i)}$ as its $\{l,i\}^{th}$ entry and $\textbf{D}$ contains $ D_i(e^{j\omega})$ as $i^{th}$ entry. 
		
		
		
\begin{figure}[!b]
			\vspace{-0.25cm}
			\centering
			\includegraphics[scale=0.65]{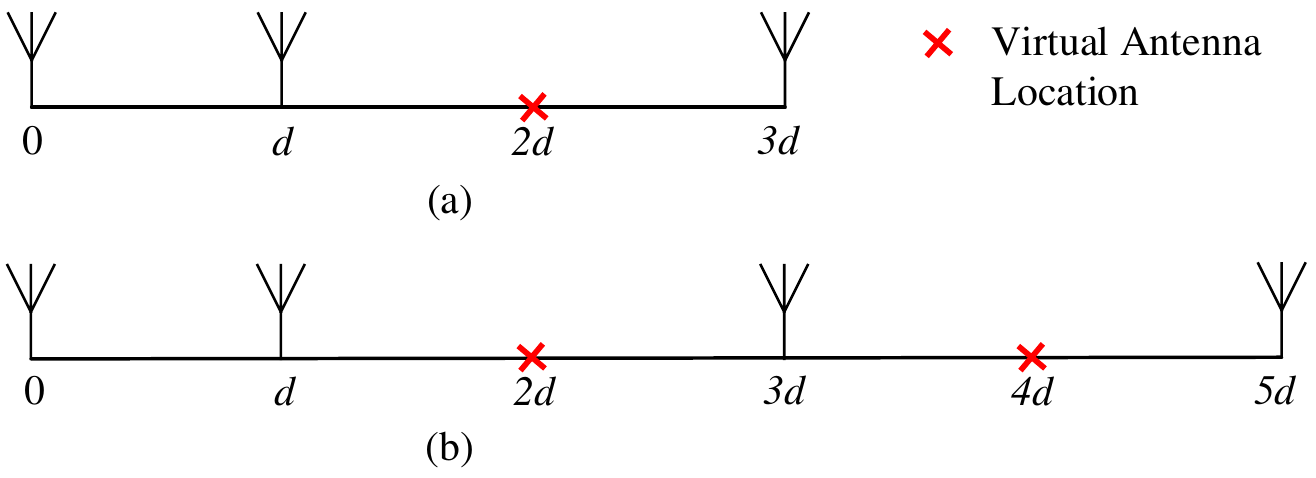}
			\caption{Sparse antenna array arrangement for (a) $L=3$ and (b) $L=4$}
			\label{sla}
		\end{figure}

		The proposed UWAS receiver uses minimum sparse ruler of length $L_s$ to design sparse antenna array. Thus, with $L$ number of physical antennas, the proposed UWAS allows DoA estimation of $L_s-1$ busy bands where $L<L_s$.
		For example, as shown in Fig.~\ref{sla}, for $L=3$ and $4$ physical antennas, the total number of actual antennas that can be utilized for DoA estimation will be $4$ and $6$, respectively \cite{sparse}. 

		For the purpose of exposition, consider $L=4$ antennas for sparse antenna arrangement. Therefore, the $\tau_l(\theta_i) = \frac{d_l}{c}cos(\theta_i)$, where $d_l = [0~d~3d~5d]$. Thus, the steering matrix \textbf{E} for sparse array will be
		\begin{align}
			\textbf{E} & = \left[ {\begin{array}{cccc}
					1 & 1 & \cdots & 1 \\
					e^{j2\pi f_1 \frac{d}{c}cos(\theta_1)} & e^{j2\pi f_2 \frac{d}{c}cos(\theta_2)} & \cdots & e^{j2\pi f_M \frac{d}{c}cos(\theta_M)}\\ 
					e^{j2\pi f_1 \frac{3d}{c}cos(\theta_1)} & e^{j2\pi f_2 \frac{3d}{c}cos(\theta_2)} & \cdots & e^{j2\pi f_M \frac{3d}{c}cos(\theta_M)}\\ 
					e^{j2\pi f_1 \frac{5d}{c}cos(\theta_1)} & e^{j2\pi f_2 \frac{5d}{c}cos(\theta_2)} & \cdots & e^{j2\pi f_M \frac{5d}{c}cos(\theta_M)}\\ 
				\end{array} } \right] 
				\label{ss_x}
			\end{align}
			
			Now, in order to estimate more DoAs than the number of antennas, $L$, the procedure similar to \cite{sparse} is followed for non-contiguous frequency bands. Hence, the auto-correlation of $\textbf{Z}_l$ will be
			\begin{align}
				\textbf{R}_{z,z} &= \int_{f\in\mathcal{B}} \textbf{Z}_{l}(e^{j\omega}) \textbf{Z}_{l}^H(e^{j\omega}) df 
				& = \textbf{E} \textbf{D} \textbf{D}^H \textbf{E}^H\\
				& = \textbf{E} \textbf{R}_{d,d} \textbf{E}^H
				\label{DoA_sparse1}
			\end{align} 
			Since the SC-FDMA signals transmitted at every frequency band are uncorrelated, $\textbf{R}_{d,d}$ is a diagonal matrix. Hence, we can apply Kronecker and Khatri-Rao properties to vectorize Eq.~\ref{DoA_sparse1}
			\begin{align}
				vec(\textbf{R}_{z,z}) &= \textbf{E}^*\otimes\textbf{E}~vec(\textbf{R}_{d,d})\\
				&=\textbf{E}^*\odot\textbf{E}~\textit{\textbf{q}}
				\label{sparse_2}
			\end{align} 
			where $\textit{\textbf{q}} = diag(\textbf{R}_{d,d})$ and $\otimes$ and $\odot$ are Kronecker and Khatri-Rao operators. The $\textbf{E}^*\odot\textbf{E}$ is a $L^2\times M$ matrix but it contains $L^2-(2L_s-1)$ redundant rows. Thus, by removing the redundant rows and re-arranging Eq.~\ref{sparse_2} in the ascending order of steering vector, we get
			\begin{equation}
				\begin{bmatrix}
					r_{-(L_s-1)} \\ \vdots \\ r_0 \\ \vdots \\ r_{L_s-1} 
				\end{bmatrix}
				= 
				\underbrace{\begin{bmatrix}
						e^{j2\pi f_1 \frac{-(L_s-1)d}{c}cos(\theta_1)} & \cdots & e^{j2\pi f_M \frac{-(L_s-1)d}{c}cos(\theta_M)}\\
						\vdots & \vdots & \vdots \\
						1 & \cdots & 1 \\
						\vdots & \vdots & \vdots \\
						e^{j2\pi f_1 \frac{(L_s-1)d}{c}cos(\theta_1)} & \cdots & e^{j2\pi f_M \frac{(L_s-1)d}{c}cos(\theta_M)}\\
					\end{bmatrix}}_{\textbf{E}_{new}}
					\textbf{\textit{q}}
				\end{equation}
				where $\textbf{E}_{new}$ is a steering matrix of size $(2L_s-1)\times M$. Now, similar to \cite{sparse}, $L_s$ vectors are generated for every $\textbf{r}_{l_s} = [r_{l_s - (L_s-1)}, \cdots , r_{l_s-1}, r_{l_s}]^T$
				$\forall~l_s = \{0,\cdots, L_s-1\}$. Subsequently, a sample average of the auto-correlation of these vectors is calculated as
				\begin{equation}
					\textbf{R}_{L_s} = \frac{1}{L_s} \sum_{l_s = 0}^{L_s-1} \textbf{r}_{l_s} \textbf{r}_{l_s}^H
					\label{R_{Ls}}
				\end{equation}
				Finally, MUSIC algorithm \cite{MUSIC} is applied on $\textbf{R}_{L_s}$ to determine the DoAs of $\beta_{busy}$ frequency bands.
				
				Please note once we estimate the status $\hat{\textbf{s}}_{\beta}$ in SS unit, the next task is to determine the carrier frequency of the user for which $\hat{\textbf{s}}_{\beta} = 1$. For  simplicity of our analysis, we  consider the carrier frequency of a SC-FDMA signal in $U_i$ frequency band is same as its center frequency i.e. $f_i = f_r + i\frac{B}{2}~\forall~i\in\{1, 2, ....,N\}$. 
				However, this assumption can be removed by applying the MUSIC algorithm on the possible sets of carrier frequencies of $\beta_{busy}$ frequency bands. 
				
\begin{figure}[t]
					\vspace{-0.2cm}
					\centering
					\includegraphics[scale=0.5]{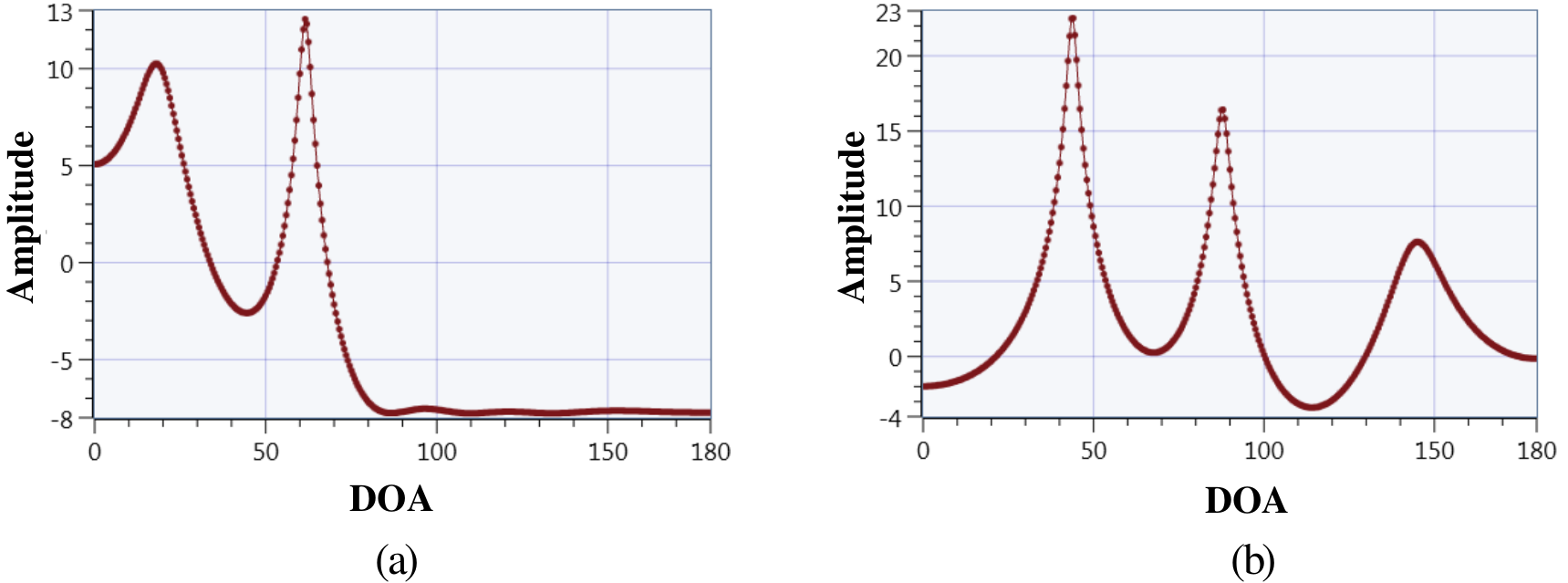}
					\caption{MUSIC spectrum (a) For two directional users and (c) For three directional users}
					\label{Music_spectrum}
					\vspace{-0.25cm}
				\end{figure}

				
				Now to determine $\theta_i~\forall ~i\in\beta_{busy}$ , we apply MUSIC algorithm. Here an over-complete steering matrix, $\textbf{E}_c$, is generated where $\theta$ varies from $0^0$ to $180^0$ with a grid size of $0.5^0$ for every $f_i$. Then the MUSIC spectrum is generated as
				\begin{equation}
					P(\theta) = \frac{1}{\textbf{e}(\theta)^H \textbf{V}_n \textbf{V}_n^H \textbf{e}(\theta)} 
				\end{equation}
				where $\textbf{e}(\theta)$ is a steering vector of $\textbf{E}_c$ for a particular $\theta$ and $\textbf{V}_n$ is the noise subspace of the auto-correlation of $\textbf{R}_{L_s}$. The peaks in the MUSIC spectrum correspond to the DoAs of transmissions present in the sensed spectrum. For example, two and three peaks in the MUSIC spectrums shown in Fig.~\ref{Music_spectrum} denote the presence of two busy bands with DoAs $18^0$ and $62^0$, and three busy bands with DoAs $42^0$, $87^0$ and $145^0$, respectively.

				\subsubsection{LDM Unit}
				To sense a large number of frequency bands, $|\beta|$ should be as high as possible.
				But as discussed in Proposition~1 \cite{was5},  for ULA antenna arrangement, the UWAS incurs sensing failure if the number of busy bands, i.e. 	$|\beta_{busy}| \geq L$.
				In case of the proposed sparse UWAS and as mentioned in Lemma~1, $|\beta_{busy}| < L_s$ for the successful sensing of  $\hat{{\textbf{s}}}_{\beta}$ and $\theta_i$ of $\beta_{busy}$ bands. Thus, there is a trade-off between $|\beta|$ and successful sensing. To overcome this trade-off, the LDM unit aims to perform the following two tasks:
				\begin{enumerate}
					\item Learning the probability statistics of the wideband spectrum. 
					\item Determining the highest possible value of $|\beta|$ and corresponding $\beta$ frequency bands for the next time slots such that sensing failure does not happen.
				\end{enumerate}

				\textbf{Lemma~1:}~If $L_s$ is the length of the minimum sparse ruler then under the noiseless condition, the perfect recovery of the occupancy status and DoA of $\beta_{busy}$ bands is possible only if
				\begin{enumerate}
					\item $K\geq L_s-1$
					\item $|| {\textbf{s}}_{\beta} ||_0 < L_s $ i.e. $|\beta_{busy}| < L_s$\\
				\end{enumerate}
				
				To perform above tasks, the following three LDM methods have been integrated in the proposed UWAS testbed.\\
				
				\noindent
				\textbf{Ideal Myopic Policy (IMP): } This policy \cite{imp} assumes the prior knowledge of spectrum statistics i.e. transition probabilities, $\textbf{p}_{uv}$ where $u,v\in\{0,1\}$ denotes \{vacant, busy\} status of frequency bands. Based on these probabilities, IMP decides $|\beta|$ and $\beta$ by maximizing the throughput as:
				
				\begin{equation}
					|\beta| = \argmax_{|\beta| \geq L_s-1} (1-\mathbb{P}(\zeta_{\beta}=1)) \sum_{i\in \beta} \psi_i(t_s)
					\label{mod_B}
				\end{equation}  
				
				\begin{equation}
					\beta = \argmax_{\beta} (1-\mathbb{P}(\zeta_{\beta}=1)) \sum_{i\in \beta} \psi_i(t_s)
					\label{B}
				\end{equation}  
				where $\zeta_{\beta}$ is the sensing failure event with $\zeta_{\beta} = 1$ denotes sensing failure, otherwise , successful sensing of $\beta$ bands.   $\mathbb{P}(\zeta_{\beta})$ is a probability of sensing failure and $\psi_i(t_s)$ is an immediate probability of vacancy of $i^{th}$ frequency band at time slot, $t_s$. These terms are defined as
				\begin{equation}
					\mathbb{P}(\zeta_{\beta} = 1) = \begin{cases}
						0 & \mbox{if}~|\beta|<  L_s\\
						1-\sum_{b=0}^{L_s-1} \mathbb{P}(\Arrowvert\hat{\textbf{s}}_{\beta}\Arrowvert_0 = b) & \mbox{otherwise}
					\end{cases}
				\end{equation}
				
				\begin{equation}
					\label{p_uv}
					{\psi_i(t_s+1) = \begin{cases}
							{p}_{10}^i, & \mbox{if}~ i\in \beta, \hat{s}_i(t_s) = 1, \zeta_{\beta} = 0\\
							{p}_{00}^i, & \mbox{if}~ i\in \beta, \hat{s}_i(t_s) = 0, \zeta_{\beta} = 0\\
							\phi_i(t_s+1), & \mbox{if}~ i\notin \beta~ \mbox{or}~ \zeta_{\beta} = 1\\
						\end{cases}}
						\vspace{-0.1cm}
					\end{equation}
					where $\phi_i(t_s+1) = (1-\psi_i(t_s)){p}_{10}^i + \psi_i(t_s){p}_{00}^i$ and $p_{uv}^i$ is the transition probability of $i^{th}$ frequency band.\\

					\noindent\textbf{Optimal LDM (OLDM):} OLDM \cite{wcl} works in two phases, i.e. exploration phase and exploitation phase where exploration is performed with $\epsilon$ probability, whereas exploitation is performed with $1-\epsilon$ probability. The exploration phase is responsible for the learning of spectrum statistics. Hence, for accurate learning, $|\beta|=L_s-1$ is considered, and all $\beta$ frequency bands are selected sequentially. Thus, allowing sufficient learning of all frequency bands. However, to maximize the throughput, the exploitation phase utilizes the learned spectrum statistics, i.e. $\hat{\textbf{p}}_{uv}$ to determine $|\beta|$ and $\beta$, according to Eq.~\ref{mod_B} and \ref{B}. Please note ${p}_{10}$ and ${p}_{00}$ in Eq.~\ref{p_uv} are replaced with $\hat{p}_{10}$ and $\hat{p}_{00}$, respectively, in OLDM.\\

					\noindent\textbf{Wideband Upper Confidence Bound (WUCB):} Similar to OLDM, WUCB \cite{was5} also learns the spectrum statistics. WUCB defines a parameter, transitional quality index (TQI) as
					\begin{equation}
						\label{tqi}
						Q_T(t_s,i) = \psi_i(t_s) + \sqrt{\frac{\delta\cdot \ln (t_s)}{U(t_s,i)}}~\forall~i\in\{1,...,N\}
					\end{equation}
					where $\delta$ is the exploration constant, and $U(t_s,i)$ indicates the number of times the $i^{th}$ band is selected till $t_s^{th}$ time slots. 
					
					WUCB initially performs sequential sensing of all frequency bands in a group of $L_s-1$ bands for once. After this, it determines the TQI of all $N$ frequency bands. Let $\alpha_1$ and $\alpha_2$ be the two vectors storing the indices of $L_s-1$ best frequency bands which have highest $Q_T(t_s)$ and $\psi(t_s)$, respectively. Now, if $\alpha_1$ and $\alpha_2$ contain same bands, then it selects $|\beta|$ and $\beta$ according to Eq.~\ref{mod_B} and \ref{tqi}, otherwise, WUCB selects $|\beta| = L_s-1$ and $\beta$ according to Eq.~\ref{tqi} to ensure successful sensing.
					\vspace{-0.1cm}
					\section{Experimental Performance and Complexity Analyses}
					\label{Sec6}

					\begin{figure}[b]
						\vspace{-0.35cm}
						\centering	\includegraphics[scale=0.4]{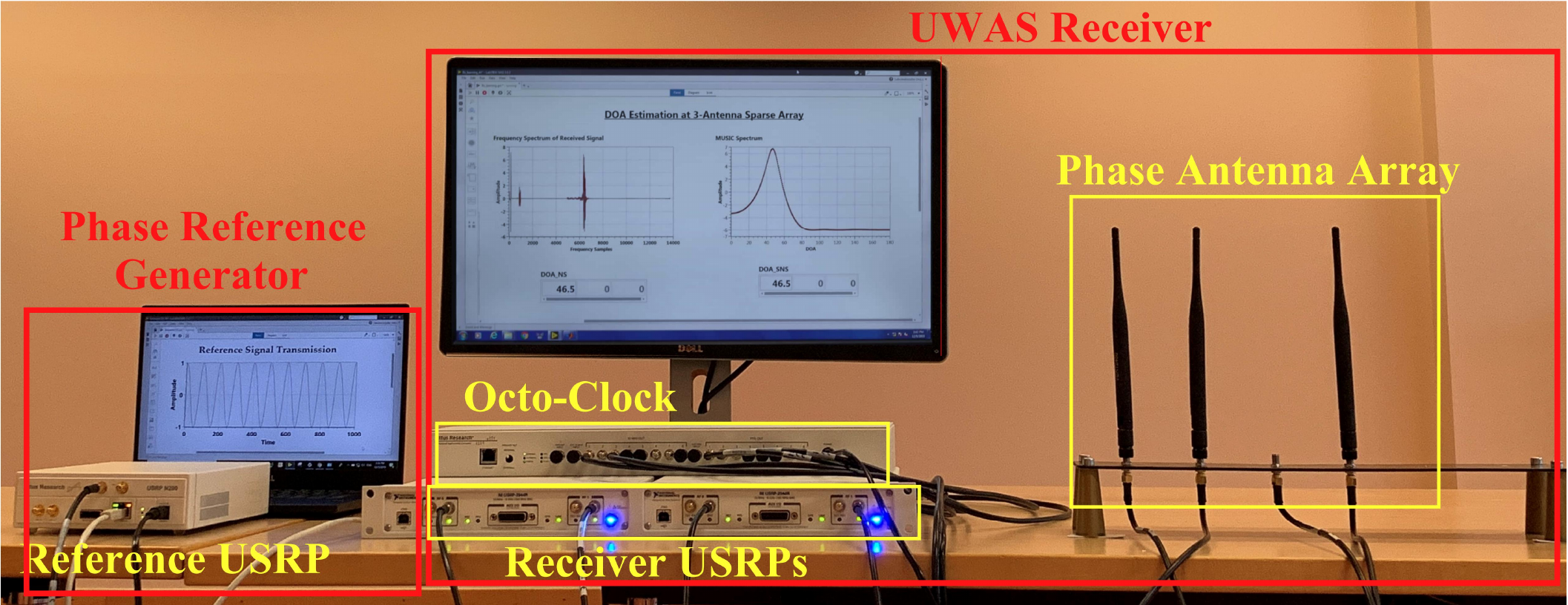}
						\caption{Proposed prototype setup consisting of phase reference generator and UWAS receiver.}
						\label{testbed}
					\end{figure}
					
					\begin{figure*}[t]
						\centering
						\includegraphics[scale = 0.35]{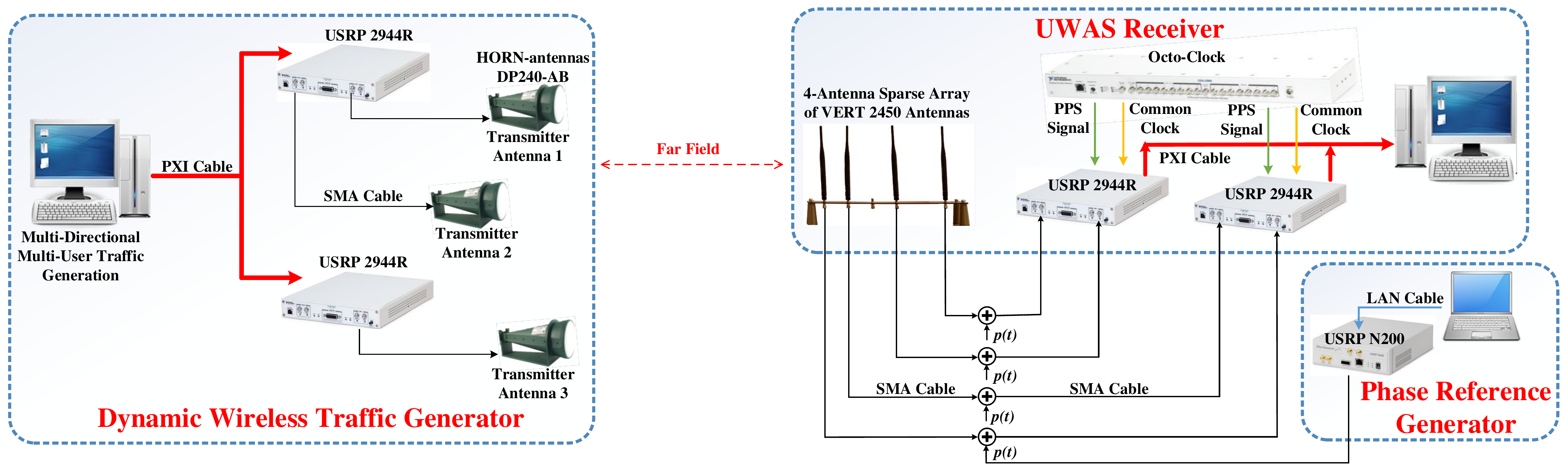}
						\caption{Graphical representation of the experimental set-up.}
						\label{Graph_testbed}
						\vspace{-0.25cm}
					\end{figure*}
					
					\	\begin{table}[htbp]
						\centering
						\caption{Transmission and reception parameters}
						\label{table}
						\renewcommand{\arraystretch}{1.3}
						\resizebox{\linewidth}{!}{
							\begin{tabular}{|c|c|c|}
								\hline
								\multicolumn{2}{|c|}{\textbf{Parameters}} & \textbf{Value}\\
								\hline
								
								\multirow{8}{*}{\textbf{Dynamic Wireless }} & $N$ & 8 \\
								\cline{2-3}
								\multirow{8}{*}{\textbf{Traffic Generator}}	 & $M$ & $3$ \\
								\cline{2-3}
								\multirow{8}{*}{\textbf{Module}} &  $B$ & $1.4~MHz$\\
								
								\cline{2-3}
								& $f_r$ & $1.8~MHz$\\
								
								\cline{2-3}
								& $f_{max}$ & $13~MHz$\\
								
								\cline{2-3}
								& Resource blocks in SC-FDMA & $6$ \\
								
								\cline{2-3}
								& {Antenna Gain} & $0~dB$\\
								
								\cline{2-3}
								& {IQ Sampling Rate} & $13~Msps$\\
								\cline{2-3}
								& {Transmission Frequency, $f_t$} & $2.4~GHz$\\
								
								\hline	
								\multirow{3}{*}{\textbf{Phase Reference}} & Carrier frequency & $200~kHz$\\
								
								\cline{2-3}
								
								\multirow{3}{*}{\textbf{Module}} & $f_{prs}$ & $400~kHz$ \\
								
								\cline{2-3}
								& {Antenna Gain} & $0~dB$\\
								
								\cline{2-3}
								& {IQ Sampling Rate} & $13~Msps$\\
								\cline{2-3}
								& {Transmission Frequency, $f_t$} & $2.4~GHz$\\
								\hline
								\multirow{7}{*}{\textbf{UWAS Receiver}}
								& $L$ & $2, 3$ and $4$\\
								\cline{2-3}
								\multirow{7}{*}{\textbf{Module}}
								& $L_s$ & $4$ and $6$\\
								
								\cline{2-3}
								
								& $K$ & $3$ and $5$\\
								
								\cline{2-3}
								
								& Common Clock & $10~MHz$\\
								
								\cline{2-3}
								
								& PPS Signal & $1$ pulse/second \\
								
								\cline{2-3}
								
								& {Antenna Gain} & $0~dB, 2~dB, 6~dB$ and $10~dB$\\
								
								\cline{2-3}
								& {IQ Sampling Rate} & $13~Msps$\\
								\cline{2-3}
								& {Reception Frequency, $f_t$} & $2.4~GHz$\\
								\hline
							\end{tabular}}
							\vspace{-0.5cm}
						\end{table}
						
						\begin{figure*}[!b]
							\vspace{-0.5cm}
							\centering
							\subfloat[]{\includegraphics[scale=0.38]{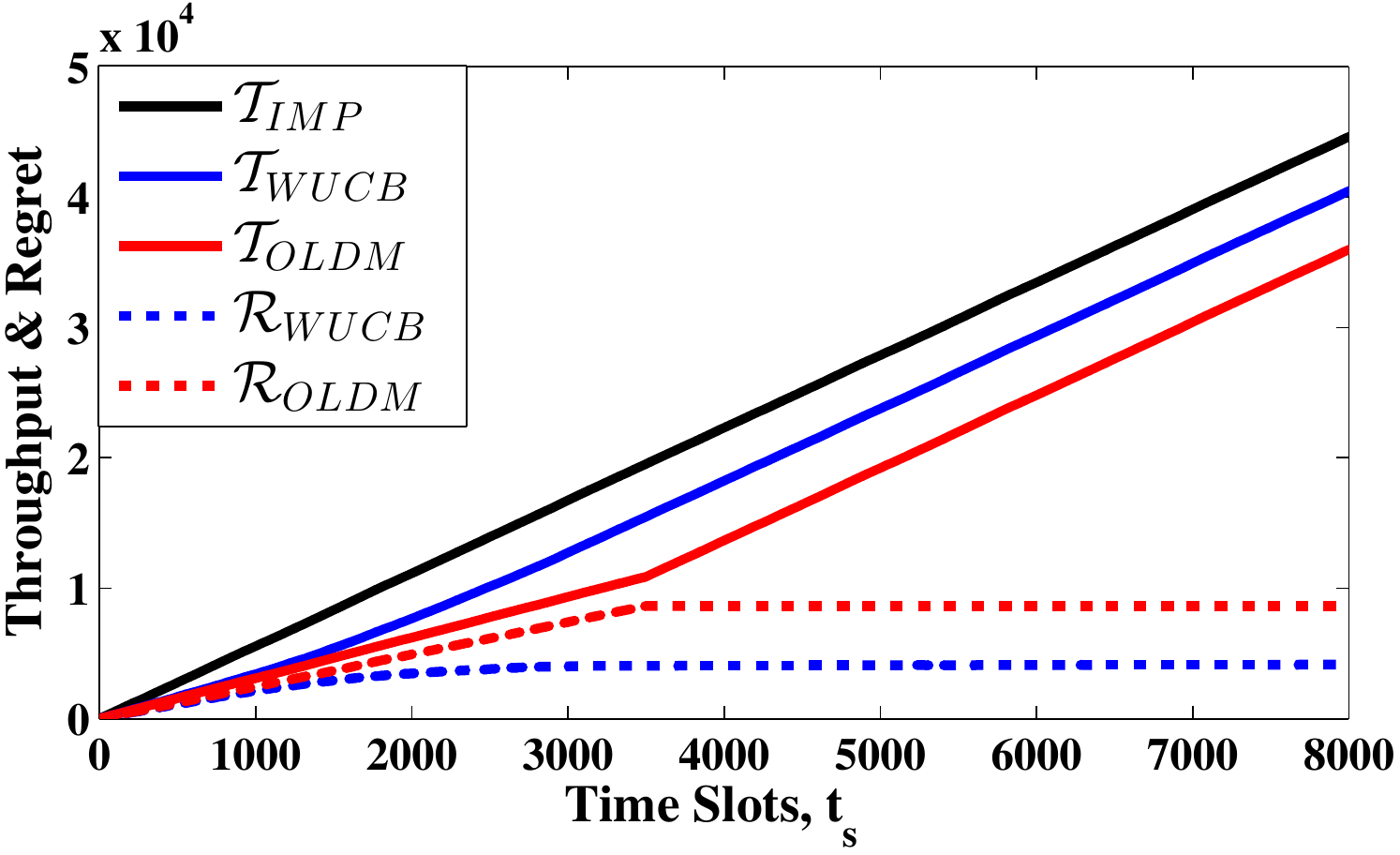}}
							\hspace{0.1cm}
							\subfloat[]{\includegraphics[scale=0.38]{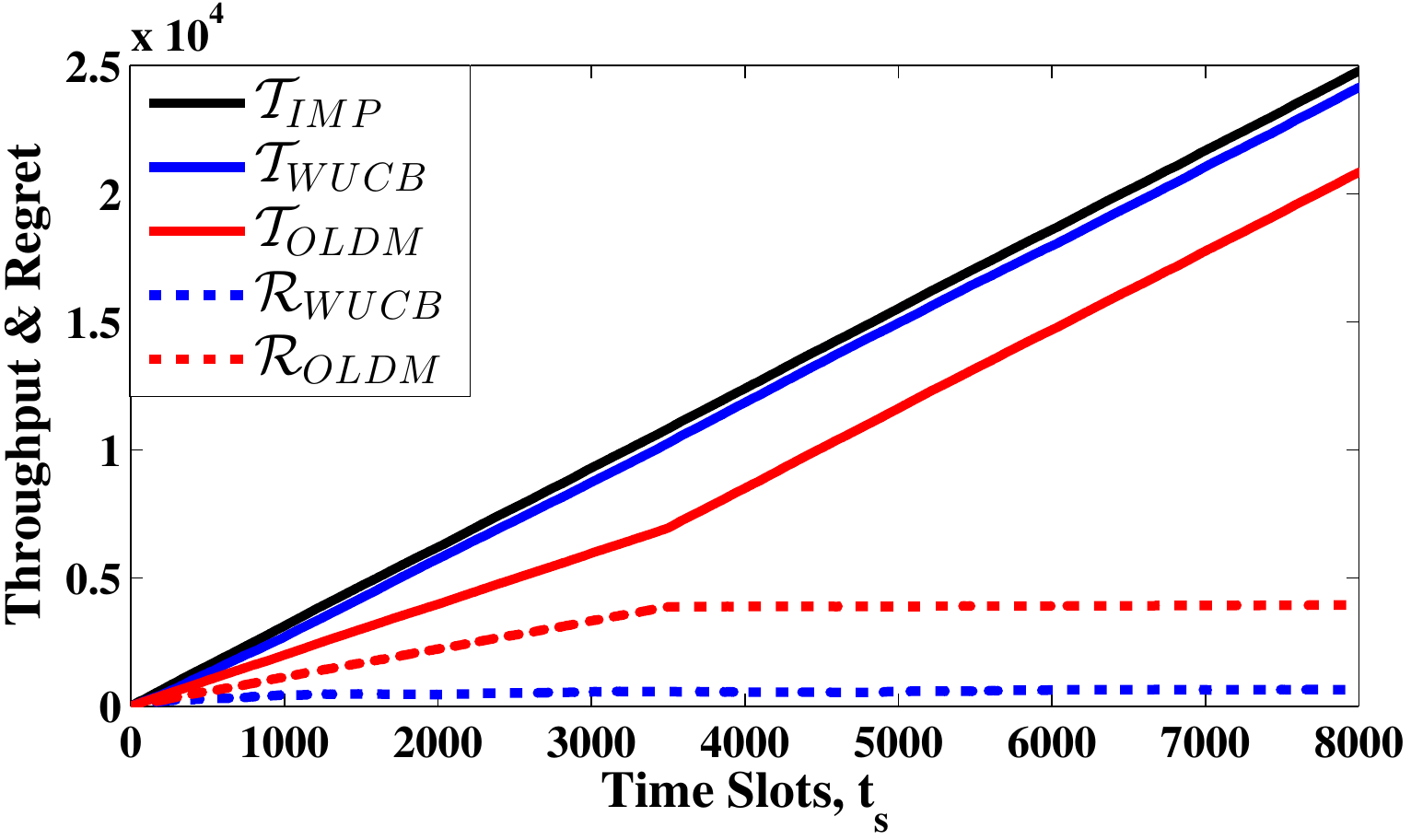}}
							\hspace{0.1cm}
							\subfloat[]{\includegraphics[scale=0.38]{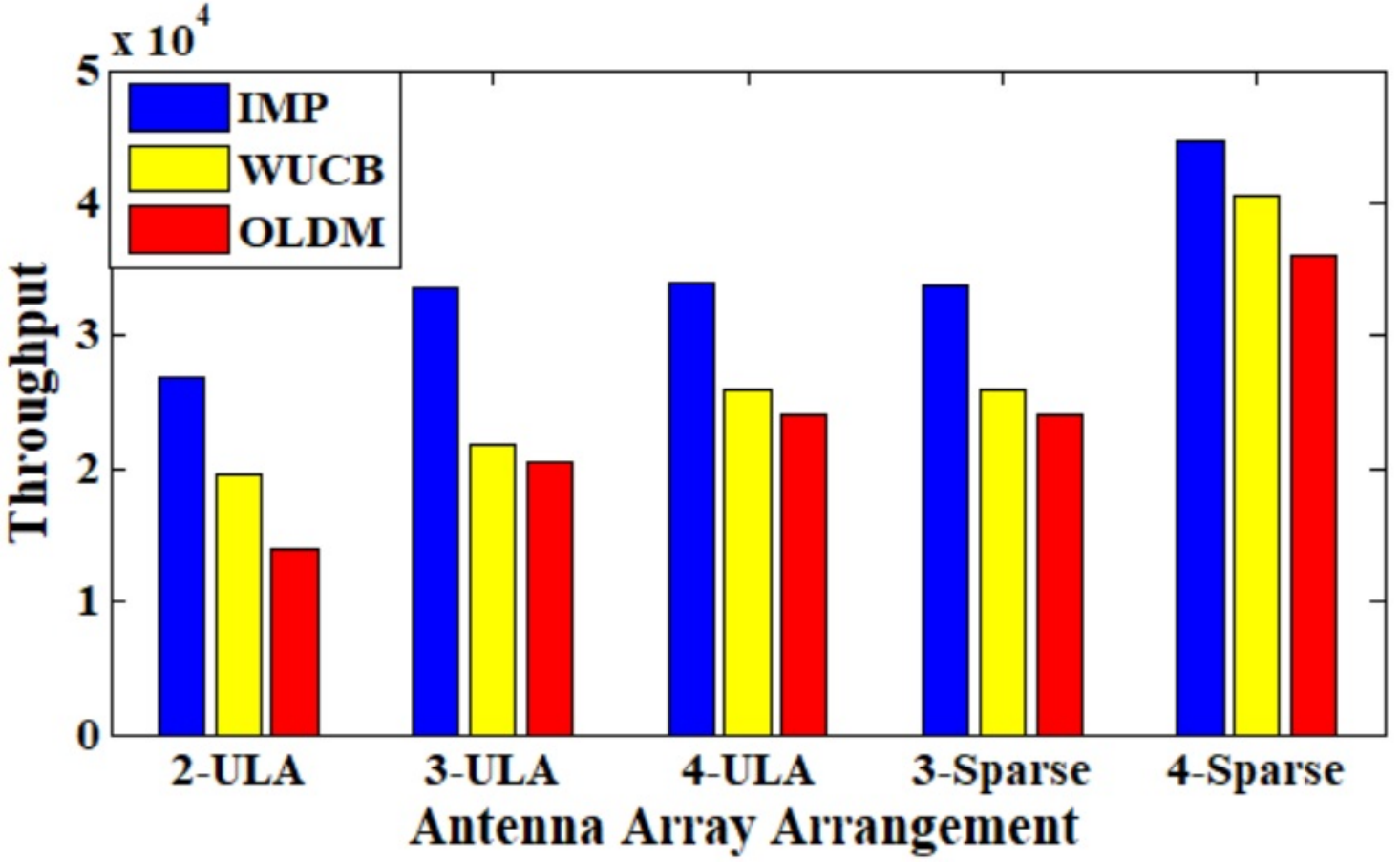}}
							\caption{Throughout, $\mathcal{T}$ and Regret, $\mathcal{R}$ achieved by the LDM methods for (a) Case~1, (b) Case~2 and (c) Various arrangements of the antenna array.}
							\vspace{-0.5cm}
							\label{LDM_results}
						\end{figure*} 		 	
						In this section, we present extensive experimental results in real-radio conditions using the proposed testbed along with the hardware complexity analysis. The prototype setup in Fig.~\ref{testbed} demonstrates the UWAS receiver and phase reference generator while dynamic wireless traffic generator is placed at distant locations to generate multi-directional multi-user traffic. As shown in Fig.~\ref{Graph_testbed}, for all the results presented in this section, the dynamic wireless traffic generator module consists of two NI-USRP 2944R with $M=3$ active transmission directions via three directional HORN-antennas DP240-AB. Since the phase reference generator module outputs common reference signal, $p(t)$ to all the AFE of the receiver, only one Ettus USRP N200 is used. The UWAS receiver module consists of two NI-USRP 2944R to provide four AFEs which are integrated with $L=4$ omni-directional VERT2450 antennas of the phase antenna array. For sparse antenna array arrangement, the antenna spacing, $d_l = ld$ where $l\in\{1,2,3\}$, are considered to be integer multiple of $d = \frac{c}{2f_t} = \frac{3\times 10^8}{2\times 2.4\times10^9}$. Various parameters of the different blocks of the proposed prototype are given in Table~\ref{table}.
						\vspace{-0.2cm}
						\subsection{Experimental Analysis} 
						To remove the reflected signals of multi-directional multi-user traffic signal, we performed the experiment in an anechoic chamber. The performance metrics used for the performance analyses are: throughput, $\mathcal{T}$, regret, $\mathcal{R}$, DoA estimation error, $\theta_{err}$ and deviation in estimated DoA, $\varDelta$. The throughput is defined as the total number of transmission opportunities in vacant spectrum and can be determined as
						\begin{equation}
							\mathcal{T} = \sum_{t_s=1}^{T_s} \lVert 1-\hat{\textbf{s}}_{\beta}\rVert_1 - \mathcal{T}_{FN} + \mathcal{T}_{\theta}
						\end{equation}
						where $T_s$ is the total number of time slots, $\mathcal{T}_{FN}$ is the throughput due to falsely detected vacant bands and $\mathcal{T}_{\theta}$ is the throughput due to angular sensing. Regret is defined as the difference between the throughput achieved by the IMP method and the throughput achieved by the other LDM methods i.e.
						\begin{equation}
							\mathcal{R}_{LDM} = \mathcal{T}_{IMP} - \mathcal{T}_{LDM}
						\end{equation}
						where OLDM and WUCB are the two LDM methods implemented in the proposed testbed. The average DoA estimation error is calculated with respect to the DoA observed at the highest receiver gain i.e.
						\vspace{-0.15cm}
						\begin{equation}
							\vspace{-0.15cm}
							\theta_{err} = \frac{1}{T_sM} \sum_{i = 1}^{T_s} \sum_{m=1}^{M}|\hat{\theta}_{g=10}-\hat{\theta}_g|
						\end{equation}  	
						where $\hat{\theta}_{g=10}$ is the estimated DoA when the receiver antenna gain is set to $10~dB$ (i.e. the maximum gain considered in the set-up) and $\hat{\theta}_{g}$ is the estimated DoA  when the receiver antenna gain is set to $g~dB$. The deviation in DoA at a gain of $g~dB$ is calculated as
						\vspace{-0.15cm}
						\begin{equation}
							\vspace{-0.15cm}
							\varDelta = max(\hat{\theta}_g)-min(\hat{\theta}_g)
						\end{equation}
						where $max(\hat{\theta}_g)$ and $min(\hat{\theta}_g)$ denotes the maximum and minimum values of $\hat{\theta}_g$ observed for the fixed antennas position.

						The performance of the UWAS testbed in terms of throughput and regret is shown in Fig.~\ref{LDM_results}. The transmitter and receiver antenna gains are set to $0~dB$ and $10~dB$, respectively. Two spectrum statistics are considered for the analyses.\\
						\textbf{Case~1}:\\ $\textbf{p}_{10}=[0.95~0.9~0.85~0.8~0.75~0.7~0.65~0.6]$\\
						$\textbf{p}_{01}=[0.05~0.1~0.15~0.2~0.25~0.3~0.35~0.4]$\\
						\textbf{Case~2}:\\
						$\textbf{p}_{10}=[0.95~0.9~0.85~0.8~0.75~0.7~0.65~0.6]$\\
						$\textbf{p}_{01}= [0.95~0.9~0.85~0.8~0.75~0.7~0.65~0.6]$

						\begin{figure*}[!t]
							\centering
							\vspace{-0.25cm}
							
							\subfloat[]{\includegraphics[scale=0.55]{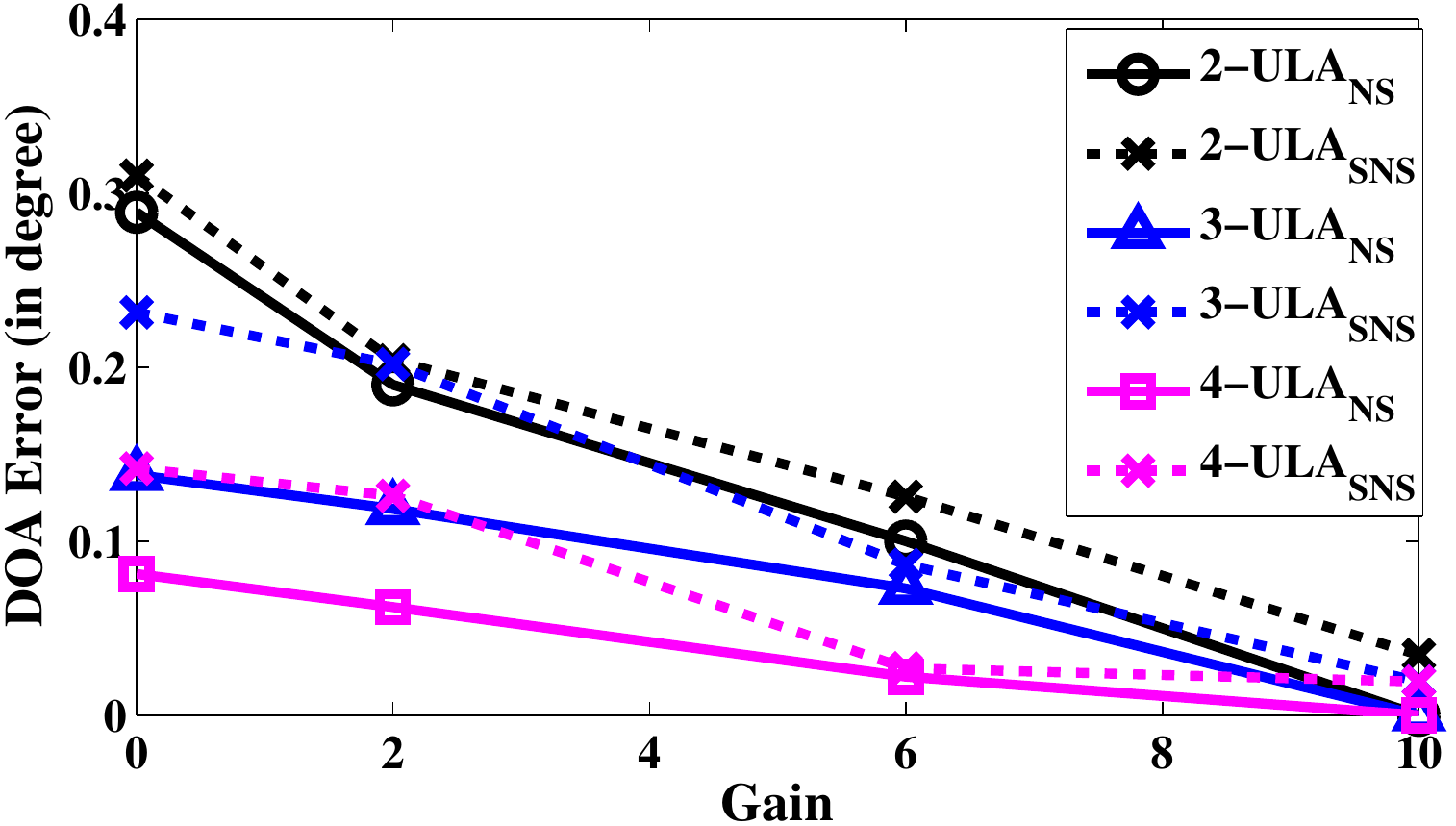}}
							\hspace{0.5cm}
							\subfloat[]{\includegraphics[scale=0.55]{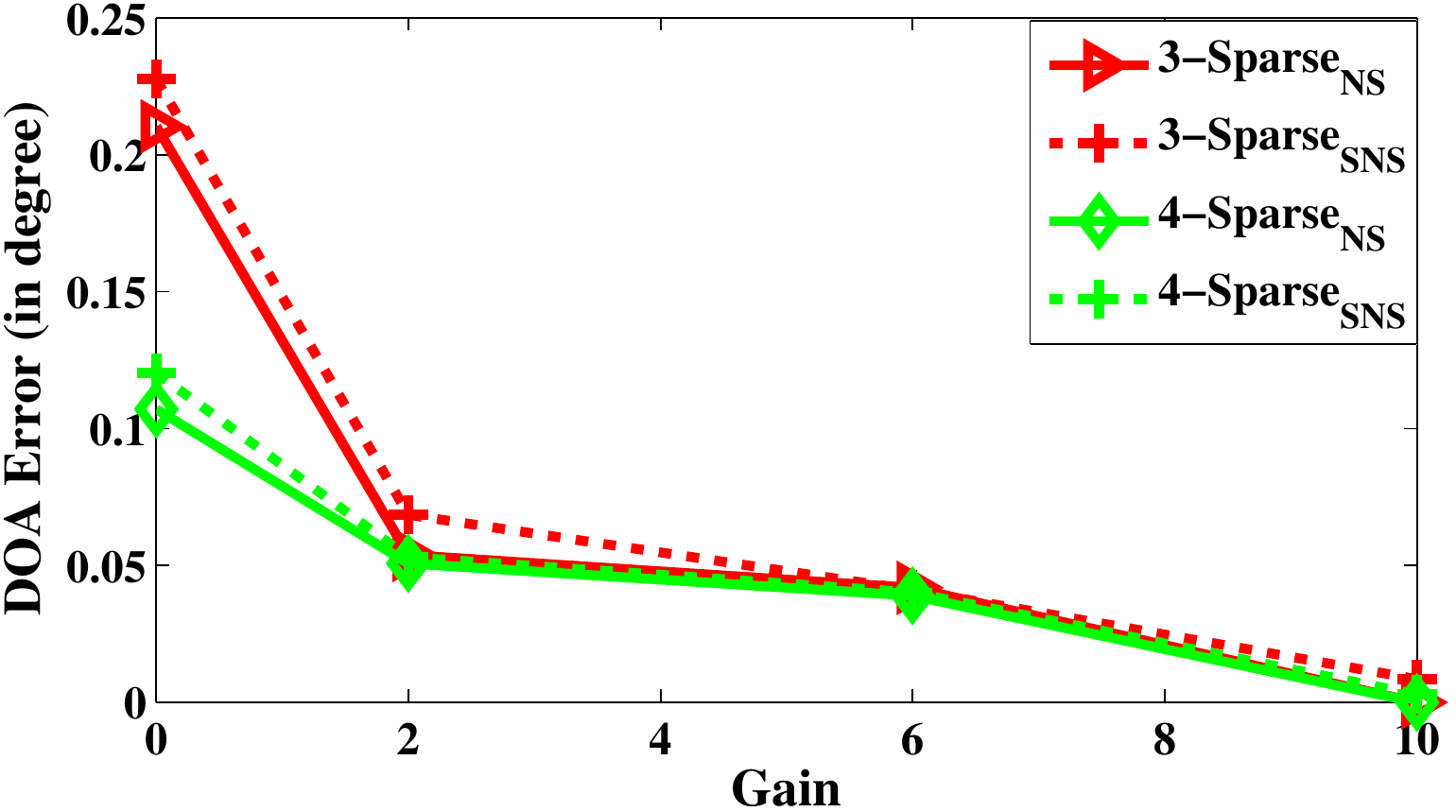}}
							
							\caption{DoA estimation error in case of single directional user signal for (a) ULA antenna arrangement and (b) Sparse antenna array arrangement.}
							\vspace{-0.25cm}
							\label{DoA_Error1}
							
						\end{figure*}

						\begin{figure*}[!t]
							\centering
							\subfloat[]{\includegraphics[scale=0.55]{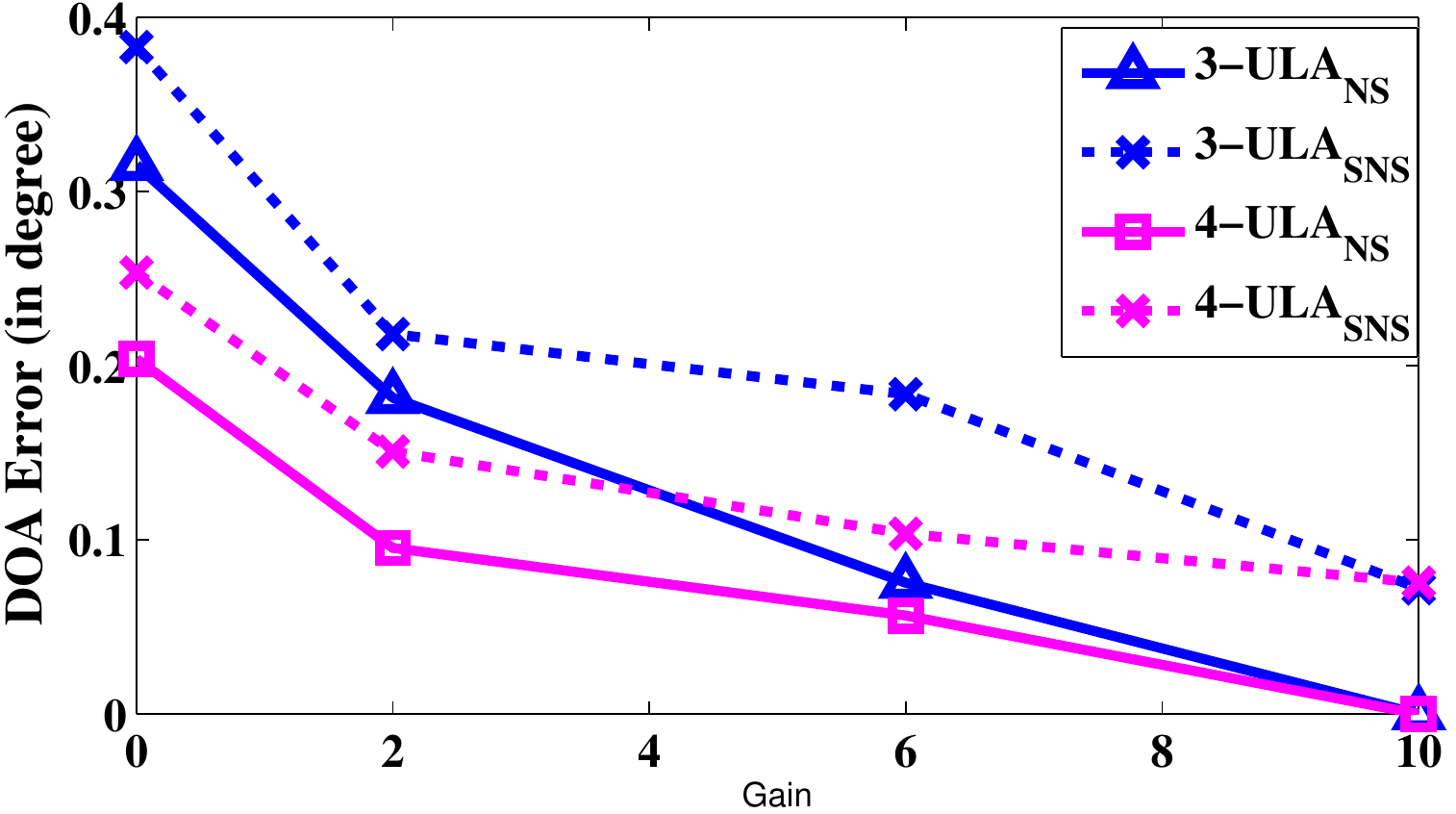}}
							\hspace{0.5cm}
							\subfloat[]{\includegraphics[scale=0.55]{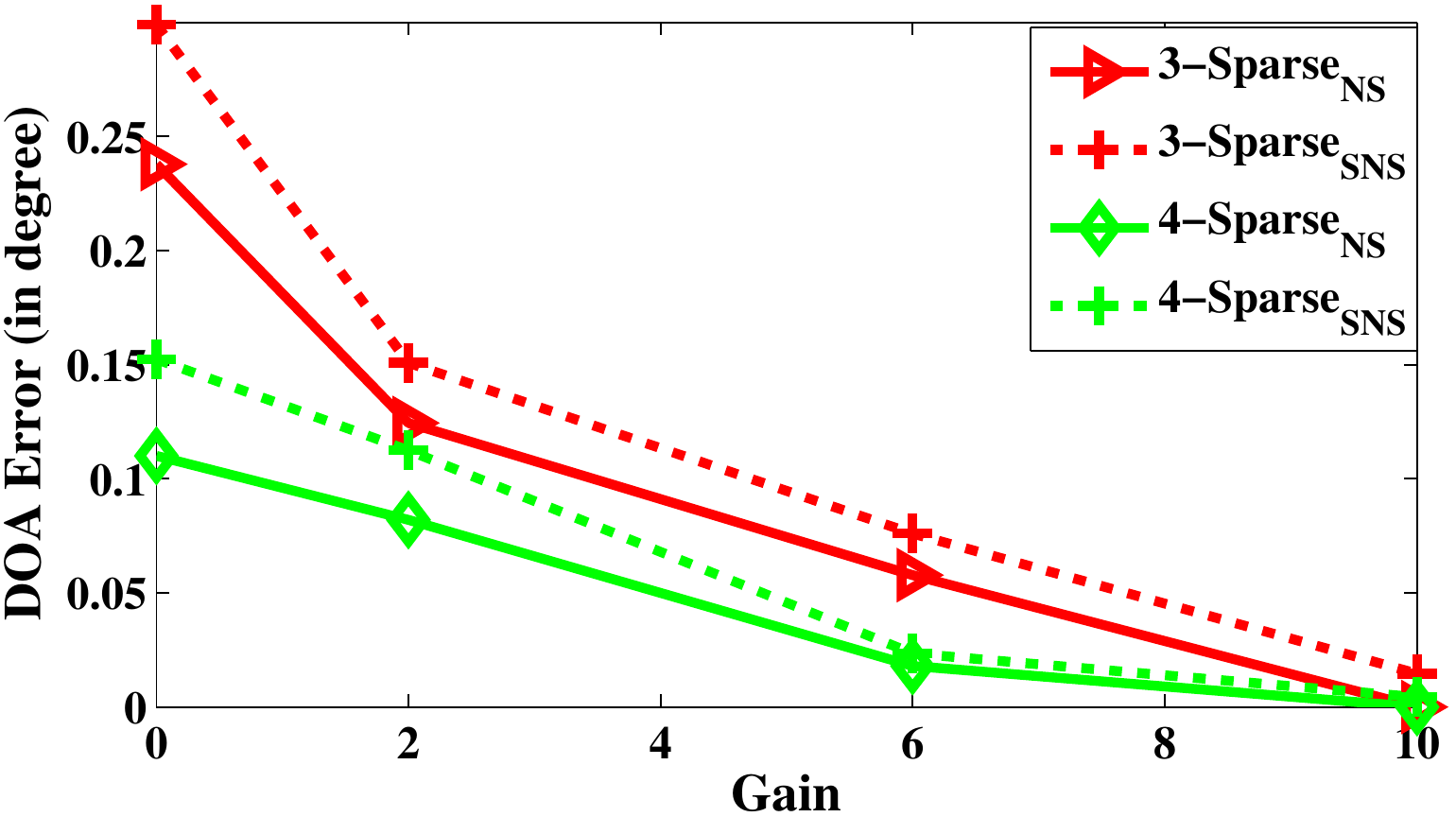}}
							
							\caption{DoA estimation error in case of two directional user signals for (a) ULA antenna arrangement and (b) Sparse antenna array arrangement}
							\vspace{-0.25cm}
							\label{DoA_Error2}
						\end{figure*}

						\begin{figure}[ht]
							\centering
							\includegraphics[scale=0.55]{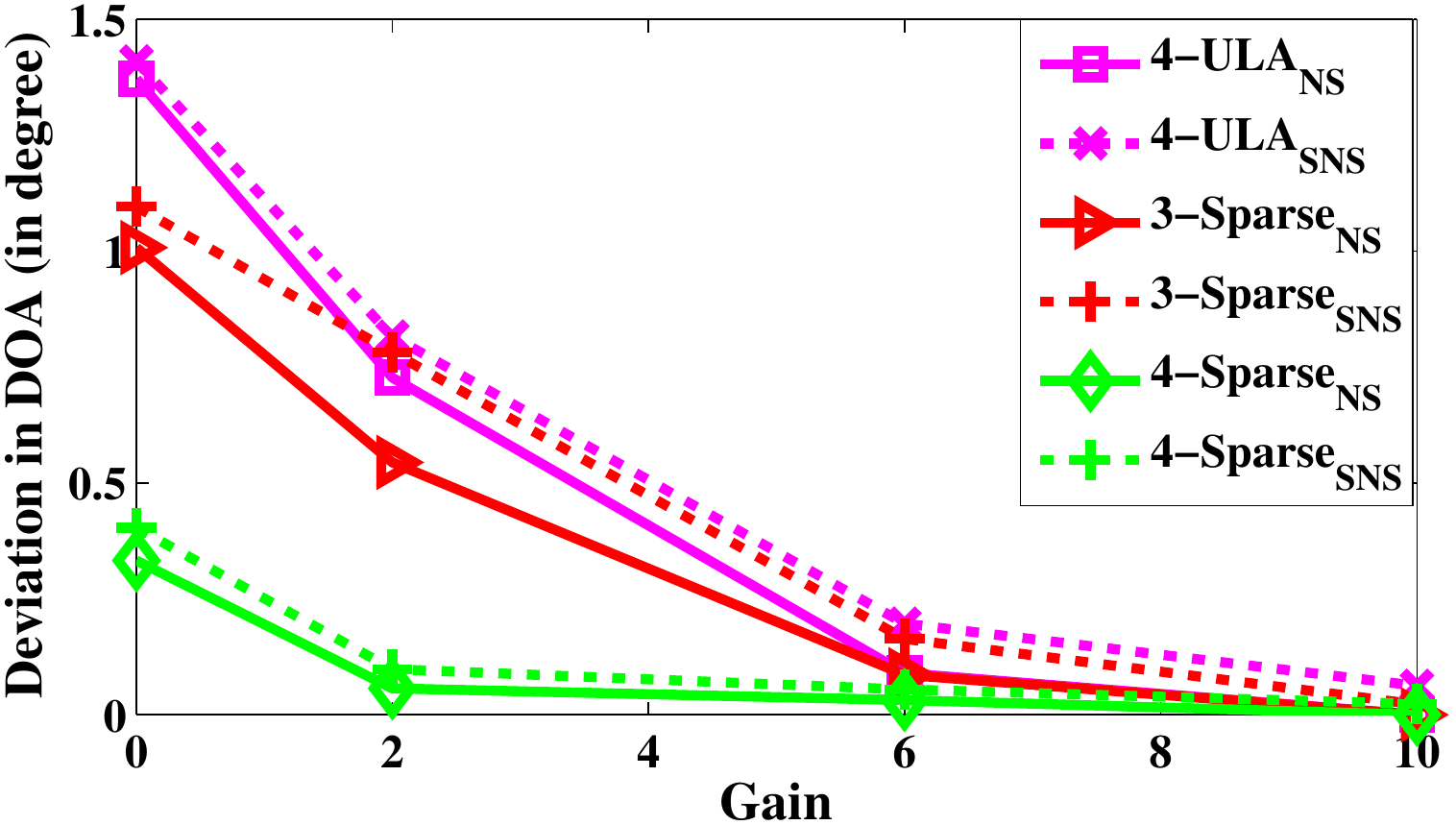}
							\caption{DoA estimation error for three directional user signals}
							\label{DoA_Error3}
							\vspace{-0.5cm}
						\end{figure}

						It can be observed from Fig.~\ref{LDM_results} that the throughput of IMP is maximum, followed by those of WUCB and OLDM. This happens because IMP has the prior knowledge of spectrum statistics, and thus it always selects the optimal value of $\beta$ and $|\beta|$. Whereas WUCB has a higher throughput than OLDM, validating the simulation results presented in \cite{was5}.  Similar observations can be verified from regret plots in Fig.~\ref{LDM_results} where instantaneous regret becomes zero (i.e. no increase in cumulative regret) after initial learning and accurate estimation of $|\beta|$. Zero instantaneous regret guarantees the convergence of  the WUCB and OLDM to the IMP, which is the desired requirement of the LDM unit. This also validates the functionality of the LDM unit in the real-radio environment compared to existing simulation-based analysis. Note that the throughput achieved by all the LDM methods is higher for Case~1 than Case~2. It occurs because the spectrum in Case~1 is more sparse and hence offers higher transmission opportunities than Case~2. This is evident from the stationary probability of vacancy, $p_0 = \frac{p_{10}}{p_{10}+p_{01}}$, which is higher in Case~1 than Case~2.

						The effect of antenna array arrangement, i.e. ULA \cite{was5} and proposed sparse antenna array based UWAS on the throughput for various LDM methods are compared in Fig.~\ref{LDM_results}(c). The analysis is done with a receiver gain of $10~dB$ for both antenna array arrangements. We considered $2$, $3$ and $4$ antenna ULA and referred them as $2$-ULA, $3$-ULA and $4$-ULA, respectively. For sparse arrangement, we considered $3$ and $4$ antenna sparse array, and referred them as $3$-Sparse and $4$-Sparse, respectively. For $3$-Sparse, antennas are placed at location $\{0,1,3\}$ whereas for $4$-Sparse, antennas are present at locations $\{0,1,3,5\}$, thereby enabling the sensing for $3$ and $5$ directional signals, respectively. It is observed that due to the increase in the number of antennas from $2$-ULA to $4$-Sparse array, the throughput of the intelligent UWAS also increases. Furthermore, since the number of possible antennas in case of $4$-ULA and $3$-Sparse are same (i.e. $4$), the throughput of all LDM methods also remain the same for both $4$-ULA and $3$-Sparse antenna arrangement.


						\begin{figure*}[!t]
							\centering
							\subfloat[]{\includegraphics[scale=0.37]{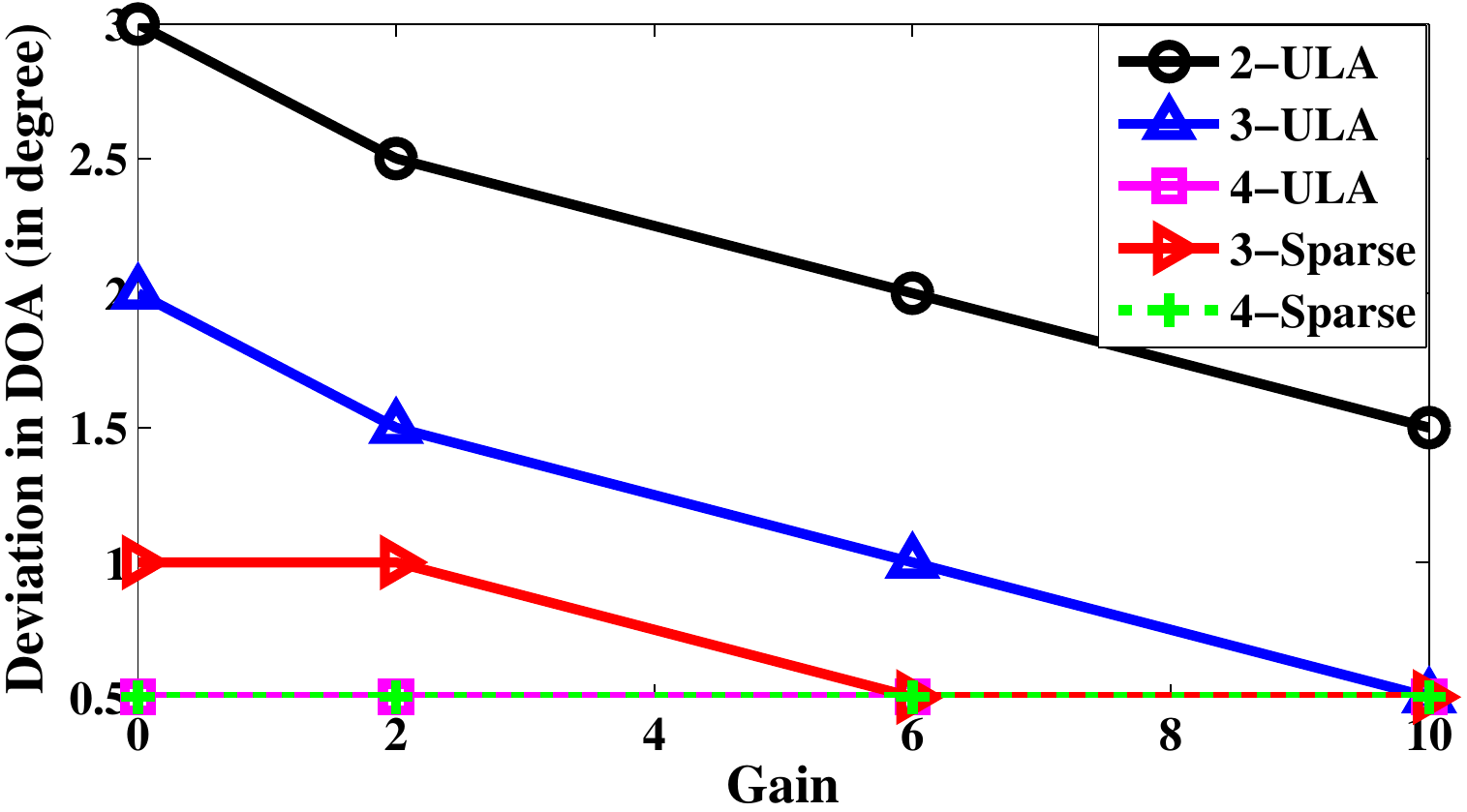}}
							\hspace{0.1cm}
							\subfloat[]{\includegraphics[scale=0.37]{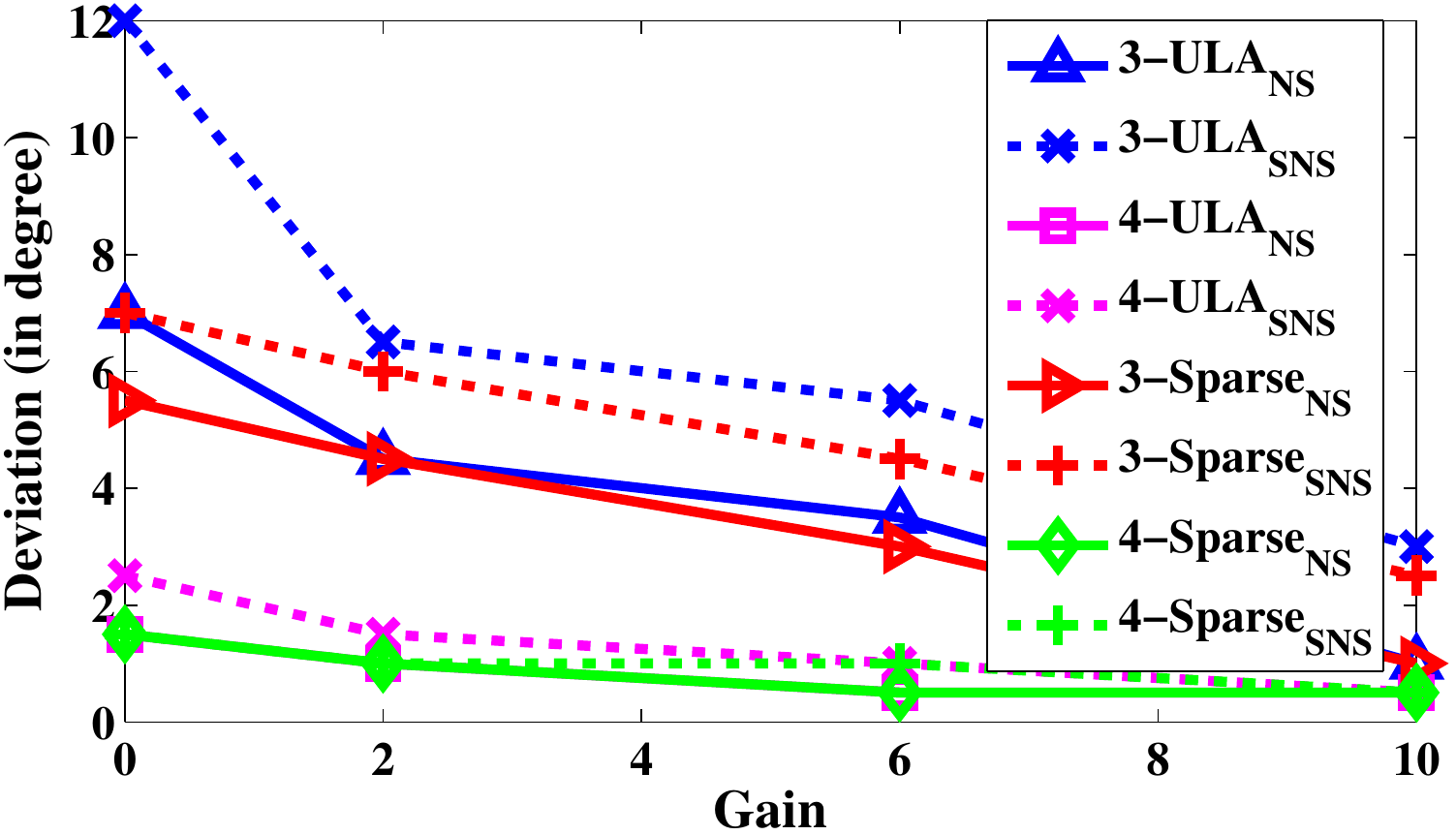}}
							\hspace{0.1cm}
							\subfloat[]{\includegraphics[scale=0.37]{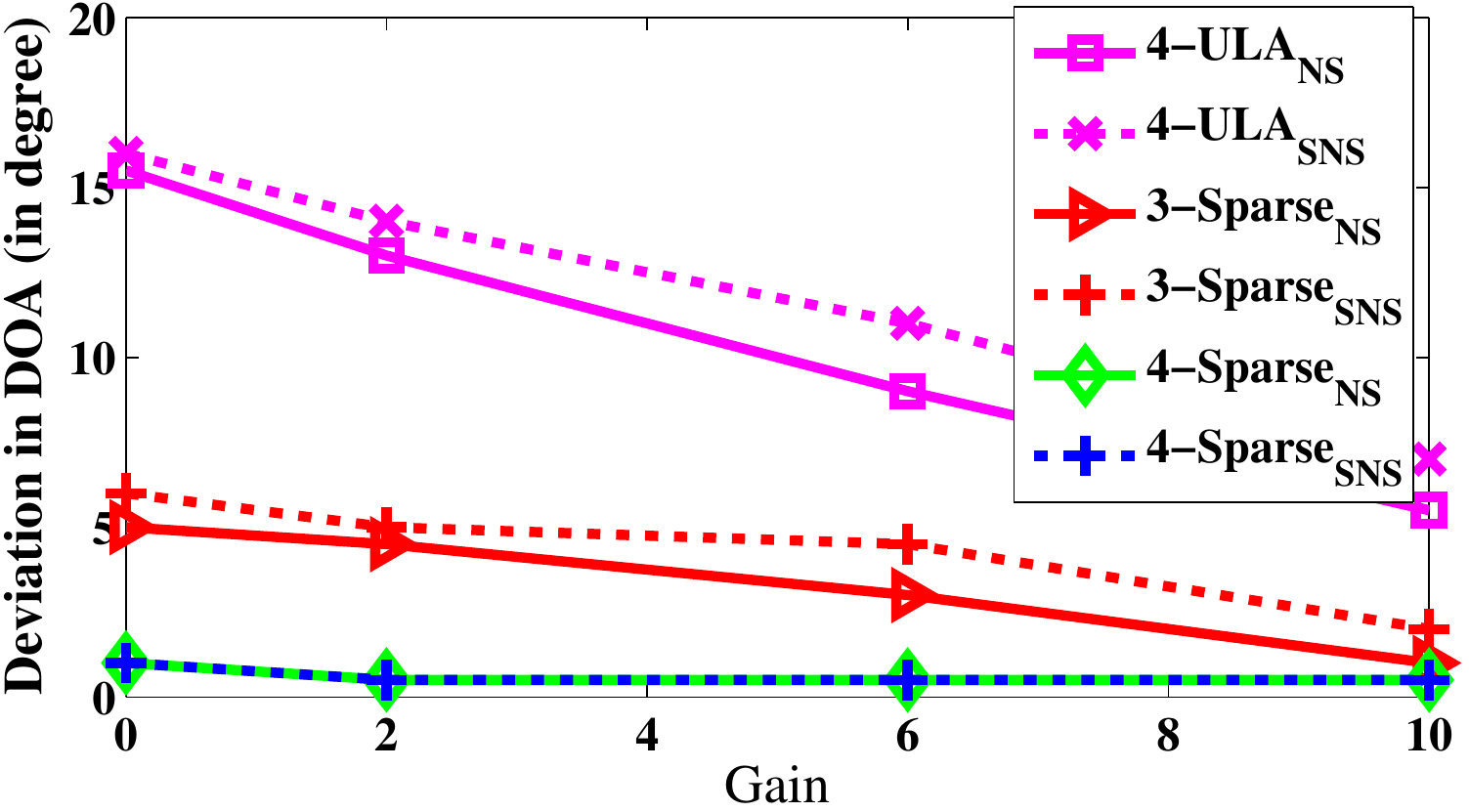}}

							\caption{ The deviation in DoA measurement for (a) One directional user signal, (b) Two directional user signals and (c) Three directional user signals}
							\vspace{-0.05cm}
							\label{T_DoA_result}
						\end{figure*}

						\begin{table*}[!t]
							\centering
							\caption{Hardware complexity comparison of different UWAS approaches}
							\label{tab2}
							\renewcommand{\arraystretch}{1.4}
							\resizebox{\textwidth}{!}{
								\begin{tabular}{|l|c|c|c|c|c|c|c|c|}
									\hline
									\multicolumn{1}{|c|}{\textbf{Characteristics}} & \textbf{\cite{was01}} &
									\textbf{\cite{was3}} & \textbf{\cite{was1}} & \textbf{\cite{was2}} & \textbf{\cite{was4}} & \textbf{\cite{was5}} & \textbf{\cite{was6}} & \textbf{Proposed UWAS} \\ 
									
									\hline
									
									\textbf{Number of Antennas} & $M+1$ & $M+1$ & $2M+1$ & $\frac{4M}{K}$ & $< M+1$ & $M+1$ & $< M+1$ & $< M+1$\\
									
									\hline
									
									\textbf{Number of ADCs} & $2M+1$ & $M+1$ & $2M+1$ & $\frac{4M+K-1}{K}$ & $<(M+1)K$ & $2M-1$ & $< 2M$ & $< 2M-1$\\
									
									\hline

									\textbf{Analog BW of ADCs} & High & High & Low & High & High & Low & Low & Low\\						\hline
									\textbf{Precise delay} & Not required & Does not require & Does not require & Not required & Requires & Does not require & Not required & Does not require\\
									\hline
									\textbf{Sensing Failure ($M\geq L$)} & Fails & Fails & Fails & Fails when $M\geq L_s$ & Fails when $M\geq L_s$ & Does not fail & Does not fail & Does not fail\\
									\hline
									\textbf{Sensing Failure ($\beta_{busy}\geq L$)} & Fails & Fails & Fails & Fails & Fails & Fails  & Fails when $\beta_{busy}\geq L_s$& Fails when $\beta_{busy}\geq L_s$\\
									\hline
									\textbf{Constraint on $f_i$} & No & No & No & No & No & Yes & Yes & Yes \\
									\hline
								\end{tabular}
							}
						\end{table*}

						Next, we compare the DoA estimation error in degrees of the UWAS receiver with ULA and sparse array for the different number of directional user signals  and sampling methods, i.e. SNS and Nyquist sampling (NS). At the transmitter, DoA of each user is randomly selected without any prior knowledge at the receiver. The DoA estimation errors for one, two and three directional user signals are shown in Fig.~\ref{DoA_Error1}-\ref{DoA_Error3}, respectively. It can be validated that the DoA estimation error decreases with an increase in the number of antennas. As the $4$-Sparse antenna arrangement creates two more virtual antennas, the DoA error is minimum for $4-$Sparse antenna arrangement. It is also validated that the DoA estimation error increases when the number of directional users increases from $1$ to $3$.  Furthermore, since the strength of the received signal increases with antenna gain, the DoA estimation error decreases significantly. Similarly, with an increase in the gain, the performance of SNS based UWAS approaches to that of NS based UWAS.

						The deviation in DoA measurement, $\varDelta$, is shown in Fig.~\ref{T_DoA_result}(a)-(c) for one, two and three directional user signals, respectively. For one user, the deviation of SNS and NS based WAS is same; hence, the deviation for only one case, i.e. NS based WAS is shown. It is observed that for one user signal, the deviation becomes zero for $4$-ULA and $4$-Sparse antenna array.  When the number of DoA sources increased from $1$ to $3$, the deviation becomes non-zero, and since $4$-Sparse creates total $6$-antennas, the deviation is minimum for the $4$-Sparse case.  Furthermore, for a given number of antennas, the deviation increases with the number of user signals. For example, for $3$-ULA arrangement, the deviation increases as $2^{0}, 7^{0}$ and $15^{0}$ when the number of user signals increases from one, two and three, respectively.

						\subsection{Hardware Complexity Analysis}			
						In Table~\ref{tab2}, the hardware complexity of the proposed UWAS is compared with that of the existing UWAS approaches \cite{was01,was1,was2,was3,was4,was5,was6} for seven different parameters. For $M$ number of users in the wideband spectrum, the number of antennas and ADCs required in all approaches is given in the first two rows. It can be observed that the proposed approach along with \cite{was3} and \cite{was2} offers a lower number of antennas and ADCs compared to other approaches \cite{was01,was1,was4,was5,was6}. This results in huge savings in the AFE, which consumes significant area and power of the wireless receiver along with limited flexibility and upgradability. Although, compared to \cite{was3} and \cite{was2}, the proposed UWAS utilizes a slightly higher number of antennas and ADCs, the analog bandwidth of ADCs used in \cite{was3} and \cite{was2} is equal to the Nyquist rate. In contrast, the analog bandwidth in our proposed UWAS is $N$ times lower. Furthermore, unlike \cite{was2}, the  proposed approach does not require precise control over analog delay in AFE which makes it possible to realize in hardware for UWAS applications. Lower analog bandwidth of ADCs along with a fewer number of antennas and ADCs makes the proposed architecture cost-efficient.
						
						Next, we consider the sensing failure, i.e. when UWAS receiver fails to digitize and characterize the spectrum, thereby leading complete loss of transmission opportunity during that time slot. As shown in the fifth row of Table~\ref{tab2}, UWAS approaches in \cite{was01,was3,was1} and \cite{was2,was4} incur failures whenever $M \geq L$ and $M\geq L_s$, respectively, i.e. whenever the number of active transmissions/users in the wideband spectrum is higher than the number of antennas. Since the wideband spectrum ranges over a few GHz, and $L$ can have limited value ranging from 1-64, the probability of sensing failure is very high in existing methods due to contiguous sensing approach. This is because the wideband spectrum may have users from other services such as narrowband IoT, WLAN along with applications in the unlicensed spectrum. Due to the augmentation of LDM with non-contiguous digitization, the sensing failure in the proposed reconfigurable SNS does not depend on the occupancy of the wideband spectrum. Instead, the sensing failure happens when  $\beta_{busy} \geq L_s$ and hence, the design of the LDM unit is critical in the proposed approach. The only limitation of the proposed UWAS method is that a user can be present only in a single frequency band and thus limits its carrier frequency. But this assumption is valid and practical as per 3GPP communication standards where carrier frequencies can take only predefined values as per the defined carrier frequency raster.
						
						

						\section{Conclusions and Future Directions}
						\label{Sec7}
						In this work, we design a prototype of the intelligent and reconfigurable ultra-wideband angular sensing (UWAS) using USRPs and LabView NXG. We demonstrated the superiority of the proposed approach over existing state-of-the-art approaches in terms of performance analysis in the real-radio environments as well as hardware complexity comparison. The proposed approach of enabling intelligence and reconfigurability in the UWAS via learning algorithms and non-contiguous sub-Nyquist sampling is novel and offers an exciting solution for next-generation wireless networks which are expected to be deployed in ultra-wideband spectrum consisting of licensed, shared and unlicensed spectrum. Further, we demonstrated the integration of sparse-antenna array with SNS which further reduces the complexity of the analog-front-end thereby making the proposed solution area, power and cost-efficient. 
						
						In the future, we would like to extend the proposed testbed for operation in millimeter-wave (mmWave) spectrum and integrate with the 5G network. The main challenge is the availability of transceivers for mmWave spectrum, and hence, in-house designs need to be explored. For the learning perspective, we would like to explore the federated learning approach where multiple base-stations learn the spectrum together, thereby improving the convergence time. With the introduction of tight integration of core and access networks in 5G, federated learning approach seems feasible and attractive solution. We are also working on the integration of the UWAS transceiver with the radar systems, thereby enabling joint radar-communication based networks which are being explored for vehicular communication systems.

					\end{document}